\documentclass[12pt,preprint]{aastex}

\usepackage{lscape}
\shortauthors{McCabe et al.}
\shorttitle{Investigating T Tauri Disk Evolution}

\begin{document}
\newcommand{\Ha}{H$\alpha$ }
\newcommand{\brg}{Br $\gamma$ }
\newcommand{\obs}{\it data}
\newcommand{\nsf}{NSFCAM}
\newcommand{\osc}{OSCIR}
\newcommand{\lw}{LWS}
\newcommand{\sig}{$\pm$\,}
\newcommand{\ulim}{$>$}
\newcommand{\av}{$A_{v}$\,}
\newcommand{\kl}{$K-L$\,}
\newcommand{\kn}{$K-N$\,}
\newcommand{\knobs}{$K-[N]$\,}
\newcommand{\nq}{$[N]-18$\,}
\newcommand{\lncolor}{$L-N$\,}
\newcommand{\sigm}{$\sigma$\,}
\newcommand{\app}{$\sim$}

\title{Investigating Disk Evolution: A High Spatial Resolution Mid-Infrared Survey of T Tauri Stars}
\author{McCabe, C.,\altaffilmark{1,2,3} Ghez, A.M.,\altaffilmark{1,4} Prato, L.,\altaffilmark{1,3,5} Duch\^{e}ne, G.\altaffilmark{1,6}  
\\ and \\ Fisher, R.S.\altaffilmark{7}, 
Telesco, C.\altaffilmark{8}} 
\altaffiltext{1}{Department of Physics \& Astronomy, University of California, Los Angeles, CA 90095-1547, mccabe@jpl.nasa.gov, ghez@astro.ucla.edu, lprato@lowell.edu}
\altaffiltext{2}{Current address: NASA Jet Propulsion Laboratory, California Institute of Technology, MS 183-900, 4800 Oak Grove Dr., Pasadena, CA 91109-8099}
\altaffiltext{3}{Visiting Astronomer at the Infrared
Telescope Facility, which is operated by the University of Hawaii
under Cooperative Agreement no. NCC 5-538 with the National
Aeronautics and Space Administration, Office of Space Science,
Planetary Astronomy Program.}
\altaffiltext{4}{Institute of Geophysics and Planetary Physics, University of California, Los Angeles, CA 90095-1565}
\altaffiltext{5}{Current address: Lowell Observatory, 1400 W. Mars Hill Rd., Flagstaff, AZ 86001}
\altaffiltext{6}{Current address: Laboratoire d'Astrophysique, Observatoire de Grenoble, Universite Joseph Fourier -BP 53, F-38041, Grenoble Cedex 9, France, Gaspard.Duchene@obs.ujf-grenoble.fr}
\altaffiltext{7}{Gemini Observatory, 670 North A'ohoku Place, Hilo, HI 96720 \\ fisher@gemini.edu}
\altaffiltext{8}{Department of Astronomy, University of Florida, Gainesville, FL 32611, telesco@astro.ufl.edu}

\begin{abstract}
We present a high spatial resolution, 10--20\,\micron\, survey of 65 T
Tauri binary stars in Taurus, Ophiuchus, and Corona Australis using
the Keck 10 m telescopes. Designed to probe the inner $\sim$1\,AU region
of the circumstellar disks around the individual stellar components in
these binary systems, this study increases the number of binaries with
spatially resolved measurements at 10\,\micron\, by a factor of
$\sim$5.  Combined with resolved near-infrared photometry and
spectroscopic accretion diagnostics, we find that $\sim$10\% of stars with a
mid-infrared excess do not appear to be accreting. In contrast to an
actively accreting disk system, these passive disks have
significantly lower near-infrared colors that are, in most cases,
consistent with photospheric emission, suggesting the presence of an
inner disk hole. In addition, there appears to be a spectral type/mass
dependence associated with the presence of a passive disk, with all
passive disks occurring around M type stars. The presence of a passive
disk does not appear to be related to the fact that these objects are
in visual binary systems; the passive disk systems span the entire
range of binary separations present in the sample and a similar
fraction of passive disks is observed in a sample of single stars. The
possibility that the passive disks are caused by the presence of an as
yet undetected companion at a small separation (0.3 -- 3 AU) is
possible for any individual system, however, it cannot account for the
spectral type dependence of the passive disk sample as a whole. We
propose that these passive disks represent a subset of T Tauri stars
that are undergoing significant disk evolution. The fraction of
observed passive disks and the observed spectral type dependence can
both be explained by models of disk evolution that include disk
photoevaporation from the central star.

\end{abstract}

\keywords{stars:pre-main sequence -- circumstellar matter -- planetary systems:protoplanetary disks -- binaries:close}

\section{Introduction}
A key component in the formation and evolution of a star and its
surrounding planetary system is believed to be the existence and
evolution of a circumstellar disk. Ample evidence for the prevalence
of these disks has been found around the young, low mass, T Tauri
stars, including excess emission above the photosphere at infrared and
longer wavelengths \citep[e.g.,][]{beckwith90, sargent91}, strong
hydrogen emission lines indicative of the accretion of disk gas onto
the central star \citep[e.g.,][]{hartmann94}, and direct images that
resolve the disks in thermal emission \citep[e.g.,][]{dutrey98,
guilloteau99} and scattered light \citep[e.g.,][]{burrows96,
stapelfeldt98} . While the formation of these disks is a natural
consequence of the star formation process, the details of how they
evolve are poorly understood. The way in which the disk material
dissipates has important ramifications not only for the
characteristics of the central star (e.g., rotation rate), but also
for the potential for planet formation. One particular challenge to
existing theories of disk evolution is the empirical evidence for two
different timescales in the inner region, which is probed by the
infrared excess.  The first is the timescale for this inner disk
region to survive as an optically thick structure and is relatively
long \citep[$\sim3\times10^{6}$ yr on average, but up to
$1\times10^{7}$ yr for some objects,][]{haisch}. In contrast, the
second is the timescale for small dust grains to be removed from this
region and appears to be significantly shorter \citep[$\sim10^5$
yrs,][]{skrutskie90, skinner91, wolkwalter, stassun01}. The short
transition timescale has been a considerable problem for models of
disk evolution \citep[e.g.,][]{kyh96, armitage99}, which generally
invoke processes that operate on the viscous timescale of the disk
($\sim 10^{7}$ yrs). While theorists have recently begun to explore
other physical processes that could give rise to such a rapid disk
evolution \citep[e.g.,][]{clarke01}, the physical processes behind
disk evolution and dispersion are still much debated.

A key step to constraining the physical mechanisms responsible for
disk evolution is the detection of young stars with disks that are in
the process of dispersing. A number of surveys have been conducted to
search for these so-called `transition objects', stars that have disk
characteristics that are intermediate between those with clear
evidence for disks and those with no disk material \citep{skrutskie90,
skinner91, wolkwalter, duvert00, stassun01}.  Most of these surveys
take advantage of the significant difference between the near- and
mid-infrared excess emission (i.e., $K[2.2\micron]-N[10\micron]$
color) from systems with and without disks, in order to establish the
evolutionary state of a system.  However, most of this work has been
done at low angular resolution, with the mid-infrared data primarily
taken from IRAS which has a beam size of 30\arcsec\, at
12\,\micron. At the distance to the nearest sites of star formation
(d=140 pc), this corresponds to 4200\,AU, within which most young
stars are observed to have stellar companions \citep[e.g.,][]{ghez93,
leinert93, simon95}. The confusion introduced by these unresolved
companions is well illustrated by the case of V773 Tau, which is one
of the two candidate evolved disks identified by Stassun et
al. (2001), based on the lack of accretion and near-infrared excess
and considerable excess at mid-infrared wavelengths and longer.  V773
Tau is a quadruple system, with 3 stellar companions within 0\farcs3
of the primary. Duch\^{e}ne et al. (2003) show that the excess seen in
V773 Tau is actually from the recently discovered optically invisible,
infrared bright, companion only 0\farcs24 away from V773 Tau A. These
infrared companions (IRCs) occur with a frequency of \app10\% in the T
Tauri population, a suspiciously similar frequency to the observed
fraction of transition objects in low spatial resolution surveys for
evolved systems. It is clear that in order to search for transition
objects around T Tauri stars, high spatial resolution observations
that resolve the individual components in these young multiple systems
are required.

In this study, we have targeted 65 binaries in the nearby star
forming regions of Taurus, Ophiuchus and Corona Australis in order to
resolve the mid-infrared emission for each individual component and
search for the presence of transition objects. A comparison of disk
properties between stars in binary systems and a sample of single
stars taken from the literature allows us to distinguish between
several possible effects, such as mass, and the presence of a
companion. The sample of T Tauri binary stars is presented in
\S\ref{c4_sample} and the observations and data reduction are discussed in
\S\ref{c4_obs}.  The observed distribution of T Tauri stars in the \kl
vs. \kn color plane is discussed in \S\ref{results}; we have
discovered a sample of T Tauri stars that have mid-infrared evidence
for disks but that do not appear to be actively accreting. These results are
discussed in \S\ref{discuss} in light of disk evolution models.

\section{The Sample \label{c4_sample}}

A sample of T Tauri binary stars was compiled from \citet{rz93},
\citet{ghez93}, \citet{leinert93}, \citet{simon95}, \citet{ghez97b}
and \citet{walter88}.  All objects in the sample are observable from
the Keck telescopes ($\delta > -40^{\circ}$), have 2MASS colors consistent with a
reddened T Tauri star (see Appendix\,\ref{reddening}), and have
separations between 0\farcs2 and 8\arcsec\, (28--1120 AU). The lower
separation bound is set by the diffraction limit at 10\,\micron\, and
the upper bound by the largest separation detectable in all
observations assuming that the primary is centered on the arrays.  The
sample is composed of binaries located within 140 pc, in the Taurus
\citep[d=140 pc;][]{elias78}, Ophiuchus \citep[d=140 pc;][]{degeus91},
and Corona Australis \citep[d=130 pc;][]{marraco} star forming
regions. Of the 70 possible targets, 65 were observed in this study
and are listed in Table \ref{sample} by star forming region and in
order of increasing separation. LkHa 332/G1, LkHa 332/G2, HBC 412, HBC
360 and NTTS 160946-1851 were not observed because of insufficient
time. The distribution of binary separations has a median
value of 0\farcs75. As a result, half of the sample stars have
separations smaller than the canonical disk size \citep[100 AU,
e.g.,][]{shu87}, and all have separations larger than the dominant
emission region of the disk at wavelengths shorter than 10\,\micron\,
($<$ 1 AU).

The majority (66\%) of the observed systems have spatially resolved
spectroscopic measurements available in the literature (see
Table\,\ref{colors}), with spectral types ranging from G2.5 -- M7.5,
and have a median type of M2.5. For these systems we define the
primary star to be the one with the earliest spectral type, i.e.,
highest mass.  For the stars that lack spectral type information, the
component that is brightest at $K$-band is assumed to be the primary.

Many of these systems also have spectroscopic measurements of \Ha
(6563\AA) or \brg (2.166\,\micron) line emission for each individual
component in the binary, allowing us to determine the accretion status
of each star.  Historically, \Ha equivalent widths have been used to
determine the presence of accretion. For the stars that have resolved
\Ha measurements, we employ the \citet{martin98} \Ha equivalent width
limits, which take into account the spectral type dependence of
intrinsic chromospheric \Ha emission to determine whether a star is
accreting or not.  For K spectral types and earlier, if the EW of \Ha
is more than 5\AA\, then the system is accreting material. For M0-M2
spectral types, the limit is 10\AA, and for M3 spectral types and
later, it is 20\AA.  The use of \brg emission lines as an accretion
diagnostic is more recent \citep[e.g.,][]{prato97,muzerolle98}.  It is
likely that some chromospheric contamination occurs, but how much has
not been determined. If there are no resolved \Ha measurements but
resolved \brg measurements are available, the presence (EW (\brg) $>$
0) or absence of a \brg line is used to define the accretion status,
independent of spectral type.  For the stars in the sample with a
measured accretion status, 78\% of them have been defined through \Ha
line emission. Altogether, of the sample with line emission
measurements, 64\% are actively accreting (CTTS) while 36\% are not
(WTTS), or are accreting at too low a level to be detectable.

Near-infrared $K$[2.2\,\micron] and $L$[3.5\,\micron] band photometry for
each component in the binary system are available in the literature
for 48 of the binary stars in the sample \citep[][and references
therein]{wg01, pgs03, woitas01b}. In contrast, only 11 of these
systems have been previously resolved at 10\,\micron\, \citep{ghez91,
ghez94b, stanke, girart04, mccabe03}.

\section{Observations and Data Reduction\label{c4_obs}}

We obtained 10\,\micron\, images of all of the 65 systems in
Table\,\ref{sample}, and 18\,\micron\, data of 26 of them, at the
W.M. Keck Telescopes.  Complementary near-infrared observations
(2--4\,\micron) of 25 systems not previously resolved in the
3-4\,\micron\, region, and of 21 additional systems in the target
sample, were made at the NASA IRTF. The goal was to obtain
near-simultaneous $K$- and $L$-band observations of each star when
possible. Combined with information available in the literature (see
\S \ref{sample}), 57 of the 65 stars in the sample have resolved
photometry in the 2--4\,\micron\, window.  Here we describe the
observations and method of data reduction for each dataset.  A log of
all observations made can be found in Table \ref{obs}, which lists the
telescope, camera, and filters used for each observing run.

\subsection{Mid-Infrared}
\subsubsection{Observations \label{mirobs}}

High angular resolution mid-infrared observations were made at the
Keck 10-m telescopes using two instruments, OSCIR, a visiting
mid-infrared (8--25\,\micron) camera \citep{telesco00}, and the Long
Wavelength Spectrometer \citep[LWS;][]{lws}, the facility mid-infrared
imaging spectrograph.  With OSCIR, which has a 128x128 Si:As blocked
impurity band (BIB) array with a plate scale of 0.062 \sig 0.001
arcsec pix$^{-1}$, an orientation of 0\fdg05 \sig 0\fdg38, and a
corresponding field of view of 7.9\arcsec\, $\times$ 7.9\arcsec, the
stars were imaged using the $N$-band ($\lambda$=10.8\,\micron,
$\Delta\lambda$ = 5.15 \micron) and IHW18 ($\lambda$=18.1\,\micron,
$\Delta\lambda$=1.6\,\micron) filters. With LWS, a 128x128 Si:As array
with a platescale of 0.081 \sig 0.002 arcsec pix$^{-1}$ and
orientation of $-3\fdg6 \pm 1^{\circ}$ (McCabe et al. 2003), wide
band-pass imaging was not possible.  Observations were therefore made
using the narrow-band filters OCLI-O ($\lambda$=8.8 \micron,
$\Delta\lambda$=0.85\,\micron), OCLI-S ($\lambda$=12.5 \micron,
$\Delta\lambda$=1.2\,\micron), and the SiC filter (11.8 \micron,
$\Delta\lambda$=2.3\,\micron) when it became available.  LWS
provides a 10\farcs24 square field of view. The plate scale and
orientation of both cameras were calibrated using observations of wide
($>$2\arcsec) binary systems.

Both instruments were used in chop-nod mode, switching between
on-source and sky frames at an average rate of 4.5 Hz. The chopper
amplitude was set to 8--10\arcsec\, such that the off-source beam lay
outside the field of view of the array. The average on-source
integration time was 2--3 min and, for all observing runs, observations
of target stars were interleaved with observations of photometric
standards, which are listed in Table\,\ref{primphot}.

\subsubsection{Data Reduction}

Observations in the mid-infrared are background dominated, with
contributions from both thermal emission from the sky, and the
telescope and dome structure. Accurate background subtraction is
therefore critical and was achieved through double differencing chop
pairs at the two nod positions.  This double difference was repeated
for each chop-nod dataset taken and the results coadded. Most of the
binary stars observed had at least one component bright enough to be
detected in the double difference image.  For these datasets, the
background subtracted nodsets were shifted and averaged using the
centroid on the detected point source. For the fainter systems which
were not detected in a single nodset, no relative position shifting
between each of the chop-nod sets was done. In general, the difference
between the final FWHM for the shifted and non-shifted frames is
small; this is consistent with the small centroid shifts detected for
the majority (87\%) of stars observed which show rms shifts of
0\farcs02 ($<$0.3 pixel).  A small constant background offset was
usually present in the averaged images. The median background was
measured in a large annulus around the primary star (typically
4\farcs8 in radius, but dependent on the binary separation), and
subtracted from the image. The inner radius of the annulus was checked
against the encircled flux of the star to ensure that the annuli
chosen were not contaminated by stellar flux.
 
\subsubsection{Astrometric and Photometric Analysis \label{phot}}

The astrometric and photometric properties of each star in the binary
system were found through fitting observed point spread functions
(PSFs) to the data, a method based on that used by \citet{mccabe02}.
We employed a two-step approach. First, aperture photometry was done by
placing a circular aperture on the centroid\footnote{On wide binaries,
the centroid was calculated for both stars. For close binaries with
large flux ratios, the centroid was calculated for the primary, whereas
for the secondary the position of peak flux was used.} of each star
observed, providing initial estimates for the stellar position and
flux ratio.  These were then used to create a model image of the
observations, using both a model Gaussian PSF of variable FWHM and a
library of observed single star PSF's from the same night as the target.
This 3$n$ parameter model for $n$ stellar components (x and y stellar
positions, relative scale of star to model PSF) was then fitted to the
data using {\em Amoeba}, a downhill simplex technique \citep{Press}.
The fitting criteria, the minimization of the sum of the residuals
$\Sigma (data-model)^{2}$, was calculated within a 20 pixel radius
circular aperture around each component. Given that the data are
background limited, with the noise level being approximately the same
for all pixels, this was equivalent to minimizing $\chi^{2}$.  The
flux of each stellar component was then measured using a 20 pixel
aperture on the individual best-fit component model, and compared to
the flux from a photometric standard. The relative fluxes are both
aperture and airmass corrected. This method provides photometry
relative to the observed standards; the absolute calibration of the
photometric standards used is described in Appendix \ref{calib}.  The
results presented in Table \ref{photres} are from the PSF fit that
provided the smallest subtraction residuals; the measurement
uncertainties for each parameter were estimated from the standard
deviation of the fits from each PSF used.  For systems in which two
stars were detected, the separation, position angle and flux ratio of
the binary are reported in Table \ref{photres}. The corresponding
uncertainties were derived from the measurement uncertainties combined
with the uncertainties in the plate scale and orientation of the
camera. If the night was photometric, the photometry of each component
also appears in Table \ref{photres}, along with the total combined
photometry.

In a number of cases, only one star in the binary system was detected.
If that source was unresolved, a model of a single PSF star was fitted
to find the best matching parameters. Determining which component is
emitting depends primarily on the separation of the binary and whether
observations resolve the binary at other wavelengths on the same
night. For example, consecutive images at 10\,\micron\, and
18\,\micron\,, where both components are seen at 10\,\micron\,, will
clarify the source of an 18\,\micron\, detection of a single
component.  If the source of the emission cannot be distinguished as
either the primary or secondary, the magnitude appears in column 9
of Table \ref{photres} and starred to indicate that the
primary/secondary designation is lacking.  If the source is extended,
PSF fitting was done assuming a binary is present. If the system
contains blended point sources that were not accurately fitted through
PSF-fitting, aperture photometry of the total (unresolved) system is
provided.  This occurred in only 1 case (LkCa 3), for which resolved
observations were available in another 10\,\micron\, filter.

\subsubsection{Filter Comparison}

Because of the changing availability of mid-infrared instrumentation and
their associated filters at the Keck telescope during our observing
runs, four filters were used to probe these systems in the 10
\micron\, region, each with slightly different central wavelengths and
bandpasses. The majority (89\%) of the systems were observed using
either the wide $N$-band ($\Delta\lambda$ \app 5 \micron) or SiC
filters ($\Delta\lambda$ = 2.3\,\micron), which are centered on the
amorphous Si emission feature. In contrast, 11\% of the systems were
observed using the narrow band 8.8 and 12.5\,\micron\, filters
($\Delta\lambda$ \app 1\,\micron), which are centered on the edge of
the silicate feature. Since the Si feature varies in strength from
object to object and in time \citep[e.g.,][]{hanner, natta, pryz03} it
is necessary to check that no large, systematic bias was introduced by
using the slightly different 8.8\,\micron, 12.5\,\micron, SiC and
$N$-band filters. 

To do this, we used the spectra of 16 typical T
Tauri stars presented in \citet{pryz03}. Our method for
the absolute calibration of the filters (see Appendix \ref{calib}) was
applied, using the relative flux spectra of \citet{pryz03}. We
linearly interpolated over the $\sim$0.7\,\micron\, gap in the spectra
at $\sim$9.3\,\micron. The relative integrated fluxes were then
measured for each of these spectra, through each of the filters used.
We found that the SiC filter magnitudes are, on average, 0.02 mag
different from $N$-band photometry, and at most 0.1 magnitude
different.  The 8.8\,\micron\, filter is on average 0.1 mag different
from $N$-band magnitudes, but can be as much as 0.5 mag different. On
the other hand, the 12.5\,\micron\, filter is on average 0.3 mag
different, and as much as 0.5 mag different; this filter is therefore
not used in quantifying colors. Most of these discrepancies are on the
order of the average measurement uncertainties in the dataset (0.11
mag for $N$-band, 0.12 mag for SiC band and 0.1 mag for
8.8\,\micron\, data). We define the index [N] as the estimate of the
10 \micron\, flux provided by the measured SiC, $N$-band and
8.8\,\micron\, fluxes, in this order of preference in the case of
multiple measurements (where the order of preference is of most used
filter to least used filter). In the rest of the paper, we directly
compare stars measured in these different filters.

\subsection{Near-Infrared \label{nirobs}}

\subsubsection{Observations}
$K$-band ($\lambda$=2.21\,\micron, $\Delta\lambda$=0.39\,\micron) and
$L$-band ($\lambda$=3.5 micron, $\Delta\lambda$=0.61\,\micron)
observations were made at the NASA IRTF 3 m telescope, using NSFCAM
\citep{nsfcam,nsfcam2}, the facility 256$\times$256 In:Sb
near-infrared camera, on 2001 May 26--28 and 2001 Nov 28--Dec 1
(UT). Two observing modes were used depending on the required angular
resolution.  For binaries with separations greater than 1\arcsec, we
used NSFCAM in direct-imaging mode, with a plate scale of 0\farcs150
\sig 0.006/pix, an orientation of 0\fdg5 \sig 1\fdg2, and a field of
view of 37\farcs9$\times$37\farcs9.  The targets were imaged in a
box-4 dither pattern with a $\sim$19\arcsec\, dither size. This
pattern was repeated 2-4 times depending on the brightness of the
stars, with exposure times varying from 0.5 -- 20 seconds per dither
position.  For binaries with separations less than 1\arcsec, speckle
imaging observations were obtained with a plate scale of 0\farcs053
\sig 0\farcs001 per pixel \citep{wg01}, with NSFCAM running in {\em
movieburst} mode, allowing a limited amount of data to be
obtained at a high duty cycle with an integration time of 0.1
seconds.  Three to four stacks of 400 such images were taken on each
object, interleaved with similar observations of known single stars
and blank sky for calibration. Photometric standards were observed in
both filters throughout each night.

\subsubsection{Data Reduction}

Each direct imaging dither pattern was reduced in the following
manner: two dither positions, with the star placed in diametrically
opposite corners, were subtracted from each other, removing the sky
background and dark current components. This difference image was then
corrected for different pixel gains using a flat field and corrected
for bad pixels.  The subtracted pairs for each star were then
registered and added together, providing two final,
averaged, images, one with a positive image of the binary and one with
a negative image. Photometry was done on both the positive and
negative image with the same PSF fitting method used for the
mid-infrared observations (see \S \ref{phot} for details). The
resulting photometry is the average of the two, with the
uncertainty estimated from the range in the measurements.

For the speckle observations, each data cube of 400 images was
reduced; sky and dark current levels were subtracted and the images
were flat-fielded and bad pixel corrected.  A power spectrum of each
image was created and divided by the power spectrum of a star that is
known to be single and point-like. The fringe pattern in the power
spectrum is a function of the binary star separation, position angle
(PA) and flux ratio. A 2D model was fitted to the fringe pattern to
measure these parameters \citep[see][for details]{ghez95}. To resolve
the 180$^{\circ}$ ambiguity on the PA, and to obtain photometry of the
combined system, the speckle stacks were also shifted and added
\citep{ghez98}, yielding an average image of the binary system.
The photometric results presented in \S \ref{results} combine the
measured flux ratio of the two components (from the power spectra
analysis) with the photometry of the combined system to get
magnitudes and fluxes for each individual component. The binary
parameters (separation, flux ratio, PA) are from the power spectra
analysis of the final, averaged, images. In all cases, quoted
uncertainties are derived from the standard deviation of results from
the individual stacks, combined with the uncertainties in plate scale
and orientation of the camera.

\section{Results \label{results}}

Measurements that produce flux ratios and/or absolute photometric
results are presented in Table \ref{photres}.  In general, objects with
[N] band flux densities greater than 10 mJy and 18\,\micron\,
fluxes greater than 30 mJy are clearly detected. We therefore took these
values as upper limits in cases where no source was detected.
Absolute photometry was obtained from our observations for 75, 93, 89,
and 24 individual components at K, L, [N] and 18\,\micron\,,
respectively.  Relative flux measurements for systems observed during
non-photometric conditions were made for 1, 4, and 3 systems at L, [N]
and 18\,\micron\, respectively.  There are nine systems\footnote{HBC
351, HBC 358, HBC 356, HBC 355, SR 12, ROXs 31, WSB 3, ROX 43, HBC
679.} in which no stars were detected in the 8--13\,\micron\, window, and
six\footnote{LkCa 3, FV Tau/c, WSB 11, WSB 26, ROX 15, DOAR 26.} in which
nothing was detected at 18\,\micron. No new components were identified
in the mid-infrared that were not already identified at some other
wavelength. Approximately 25\% of the binary systems have inverted
flux ratios at wavelengths longer than 2\,\micron ; the
brightest component at 2\,\micron\, is not necessarily the brightest
at either 3.5\,\micron\, or 10\,\micron\,, highlighting the potential
pitfalls in assigning long wavelength, low spatial resolution
measurements to the component that dominates at shorter wavelengths.
These results increase the number of systems with spatially resolved
measurements at 10\,\micron\, by a factor of \app\,5.

We construct a final sample of 71 components in 35 systems that have
high spatial resolution K, L and [N] band measurements; these are
listed in Table \ref{colors}. All but 4 of these 35 systems also have
spatially resolved spectroscopic information, providing spectral type
and accretion diagnostics for each stellar component. Measurements
used in this final sample are first taken from our results and then
supplemented with other work. Specifically, we have added 2
measurements from the literature at mid-infrared wavelengths 
\citep[DoAr 24E and HK Tau, from][]{stanke, mccabe03}, and approximately half of the
measurements at near-infrared wavelengths (see Table \ref{colors} for
references).

For our final sample, dereddened \kl vs. \knobs and \knobs vs. \nq
\,colors are presented in Figure\,\ref{colorplane_dered}, with the
symbol type representing whether or not the star is actively
accreting. The de-reddened colors for each component are also listed
in Table \ref{colors}, along with the visual extinction for the system
(inferred using a method detailed in Appendix\,\ref{reddening}), the
spectral type when known, the spectroscopic accretion measure used
(i.e., \Ha or \brg) and the equivalent width (EW) of \Ha emission when
applicable. In both the \knobs and \nq color indices the stars are
well separated into the three groups originally identified by
\citet{lada87}, corresponding to photospheric emission (Class III),
disk emission (Class II), and embedded protostellar emission (Class
I).  Here we adopt the following mid-infrared color limits to
delineate these groups:

$~~~~~~~~~~~~~~~~~~~~~~~~~~~~~~~~~~~~~\mathrm{Class~\,I:}~~ K-[N] > 4.5  \\ $
$~~~~~~~~~~~~~~~~~~~~~~~~~~~~~~~~~~~~~~~~~~~~\mathrm{Class~\,II:}~~ 1.75 < K-[N] < 4.5  \\$ 
$~~~~~~~~~~~~~~~~~~~~~~~~~~~~~~~~~~~~~~~~~~~~\mathrm{Class~\,III:}~~ K-[N] < 1  \\$

\noindent These correspond to the \citet{wilking01} spectral index
limits, with three modifiations: we include the flat spectrum class
defined by Wilking et al. in our class I definition, we use a class II
lower limit of \knobs = 1.75 rather than \knobs=2, and we adopt a
class III cutoff that is intermediate between that suggested by
Wilking et al. ($K-N$ = 2) and Kenyon \& Hartmann (1995; \knobs \app
0.5).  The class III color limit therefore incorporates the
photospheric contribution from even the lowest mass star in our sample.

Within our sample, we have detected 3 class I sources, 58 class II
sources and 10 class III sources. Since our sample was selected for T
Tauri stars, we did not expect to detect many class I stars.  All
three class I stars (T Tau B, FV Tau/c B and DoAr 24E B) are
secondaries to class II primaries and are previously identified
infrared companions (IRCs) or high accretion stars
\citep[][]{koresko97, wg01}.  At 18\,\micron\, we have detected 2 class
I sources, 19 class II sources and 2 class III sources. We lack the
sensitivity at this wavelength to detect the majority of the WTTS in our dataset.

While the distinction between class II and class III stars remains
clear in the \knobs index in Figure\,\ref{colorplane_dered}, the \kl
index shows some overlap.  Using the \citet{kh95} cutoffs, we would
expect class III stars, with only photospheric emission, to have \kl
$\lesssim$0.4 and the boundary between class II and class I to be at \kl
$\sim$ 1.5. However, several class II objects, with clear \knobs
excesses, have \kl colors consistent with class III objects. One
possibility is that this subset of our sample has disks with inner
holes in their dust distributions.

The possibility of inner disk holes becomes even more intriguing when
the accretion diagnotistics are considered. Of the 58 class II stars,
52 have accretion information.  While the majority of stars with IR
excess (class II) are also accreting (CTT), we find that 11\% of the
observed class II stars with accretion diagnostics do not appear to be
currently accreting. We define these objects to be passive disks; the
star is surrounded by significant amounts of dust, most likely in the
form of a disk, but material does not appear to be actively accreting
onto the photosphere. These objects are highlighted in
Table\,\ref{colors} with ``passive'' in the notes column and are
discussed in more detail in Appendix\,\ref{passive_obj}. The majority
(66\%) of the non-accreting, passive disks also have \kl colors that
suggest the presence of an inner disk hole.  A K-S test comparing the
distribution of colors between the passive disks and class II-CTTS
sample show that there is only a 0.9\% probability (a 2.6\,\sigm
result) that they are sampling the same distribution in \kl. On the
other hand, there is no large difference in the distributions of
\knobs between class II-CTTS and the passive disks (8\%, or
1.8\,\sigm). The passive disks appear to be associated with the inner
disk hole population.

There is a tendency (5/6) for the passive disks to reside in mixed
systems\footnote{One passive disk, WSB 28, is in a WTTS-WTTS pair},
where a WTTS is paired with a CTTS. There is also a tendency for the
passive disk to be present around the secondary, the lowest mass
component in the binary system. Only one passive disk is found around
a primary and that is FV Tau/c A. The stellar companions to the
passive disk objects appear to be normal CTTS or WTTS based on their
\kl and \knobs color indices. Because the sample of passive disks is
small, neither tendency is statistically significant. 

The passive disks are in binary systems that span almost the entire
range of separations in the sample, from 0\farcs3 to 5\arcsec. The
presence of a companion star in the range of binary separations probed
here does not appear to affect whether a disk is passive or
not. Indeed, the presence of a passive disk does not appear to be
restricted to stars in binary systems. A comparison sample of known
single T Tauri stars with near-infrared and mid-infrared colors
available in the literature is shown in Figure\,\ref{colorsingles}
(see Appendix\,\ref{app_singles} for sample details). We now know less
about the mid-infrared properties of single T Tauri stars than we do
about those in multiple systems; we therefore have a limited sample
size for this comparison and the majority of the single star
mid-infrared colors are based on low spatial resolution IRAS
measurements.  With these caveats in mind, we find a similar fraction
of passive disks systems (1/20, or 5\%) around single stars. Whatever
phenomenon is responsible for the passive disks, it does not appear to
be related to the presence of a wide separation ($>$30 AU) stellar
companion.

A closer examination of the stellar properties of the passive disk
systems in our binary sample reveals that this phenomenon appears to
be restricted to stars with spectral types of M2.5 and later. Dividing
the class II stars into two roughly equal subsets based on whether the
spectral type is earlier or later than M2.5, we find 0 out of 28
passive disks in the early spectral type set, and 6 out of 22 in the
late spectral type set.  The distribution of spectral types for the
sample of normal class II-CTTS stars is compared to those of passive
disks in Figure\,\ref{hist_spty}, which combines both the single star
sample and binary star sample together. In this combined sample we
find 0/26 passive disks for objects with spectral types earlier than
M0, and 7/46 (15\%) for objects with a spectral type of M0 or later. A
K-S test on the two spectral type distributions shown in
Figure\,\ref{hist_spty} finds a 0.9\% probability that these samples
are drawn from the same parent distribution.  Since later type stars
are more chromospherically active, the \Ha threshold increases with
spectral type. Our correction for this employs a coarse threshold
approximation for an inherently continuous distribution of upper limit
chromospheric \Ha equivalent widths as a function of spectral type
(see \S\,\ref{c4_sample}), raising the possibility that the
non-accretion status of these passive disks may be mis-classified. We
have checked the passive disks which have \Ha measurements to see how
close to the limit the \Ha level is.  Although FQ Tau B is a
borderline CTTS/WTTS case, LkHa 332 B, FV Tau/c A and FX Tau B are
unambiguous. These results suggest that the presence of a passive disk
with an inner disk hole appears to be dependent on the spectral type,
and hence mass, of the central star.

\section{Discussion \label{discuss}}

Approximately 10\% of all class II T Tauri stars, with significant
evidence for a circumstellar disk at 10\,\micron, do not appear to be
currently accreting. Previous work has shown that the accretion
flow in T Tauri stars is quite variable \citep[e.g.,][]{smith,
alencar, alencar02, littlefair}, suggesting that a mis-identification
of WTT/CTT status can occasionally occur. If what we are observing is
entirely the result of a stochastic accretion process, then the sample of
observed passive disk stars suggests that, on average, class II stars
spend approximately 10\% of their CTT lifetime in a non-accreting
state. As we point out in Appendix\,\ref{passive_obj}, a couple of the
candidate passive disk objects may in fact be undergoing stochastic
accretion. However, a variable accretion flow cannot account for the
observed link between a non-accreting disk and the presence of an
inner disk hole, or the observed spectral type/mass dependence. In the
following subsections we explore two mechanisms that can reproduce the
observed inner disk hole in the passive disk systems: (a) the presence
of a nearby, as yet unresolved companion (either stellar or
substellar/planetary in mass, \S\,\ref{companions}), or (b) disk
evolution (\S\,\ref{evolution}).

\subsection{Evidence for unresolved companions?\label{companions}}

In principle, the presence of a nearby companion could reproduce the
observed inner disk hole and, potentially, the lack of significant
accretion through star-disk or planet-disk interactions. Such
companions could not have been imaged around the passive disk systems
in our sample as they would have been too close to resolve, and radial
velocity monitoring of these objects has, to the best of our
knowledge, not been done. In systems with a stellar companion (with
mass ratios $\lesssim$ 10), resonances between the orbit of the star
(with semi-major axis $a$) and that of the dust in the disk will
dynamically clear out material up to 2-3$a$, depending on the
eccentricity of the orbit \citep{Artymowicz94}. For planetary mass
companions it is generally assumed that the disk clearing occurs
essentially at the apoapstron of the planets orbit
\citep[e.g.,][]{quillen04}. Because the near-infrared and mid-infrared
excesses arise from the inner few AU of the disk, only systems with
companions at separations on order of 1\,AU could lead to the absence
of such excesses. Consistent with these predictions, observed
spectroscopic binaries with separations less than $\sim$0.3\,AU, do
not show the characteristics of the passive disk systems; they have
\kl excesses and accretion can still occur across the disk gap (e.g.,
DQ Tau, Mathieu et al. 1997; AK Sco, Alencar et al. 2003; and NTTS
162814-2427, Jensen \& Mathieu 1997), whereas there are two systems
with either known or putative companions at separations on the order
of a few\,AU that would be classified by us as passive disks; HD 98800
B (K5 combined spectral type), which has a {\it known} stellar
companion at an apoapstron distance of 1.6\,AU \citep[e.g.,][]{Koerner00,
Prato01}, and CoKu Tau 4 (an M1.5 star), whose SED has been modelled
assuming an {\it undetected} companion with M $> 0.1\, M_{J}$ at
10\,AU \citep{quillen04, Dalessio04}. Thus, companions with separations
on order of 1\,AU are capable of mimicking the color and accretion
properties of the passive disk systems.

While a stellar or substellar companion a few AU from the star could
be responsible for the observed lack of accretion and presence of an
inner disk hole, companions are unlikely to be the explanation for all
the observed passive disks.  The period distribution for stellar
companions to M type dwarfs (for separations $\lesssim$ 10 AU) is very
similar to that around G and K type dwarfs, with approximately 10\% of
the stars having a companion in the 0.3 -- 3\,AU separation range
\citep[][]{dm91, marchal03}. If all the passive disks were caused by a
stellar companion in this separation range, we would be equally likely
to find passive disks around K type stars as M type stars in our
sample, which is not the case. The observed mass dependence also
suggests that {\em sub}stellar or planetary companions are not the
probable cause. As \citet{butler04} show, planets with separations
$\lesssim$1\,AU are $\sim$5 times less likely to occur around M type
stars than around G or K type stars.  Therefore the fact that all the
passive disks occur solely around M type T Tauri stars makes the
presence of a companion, either stellar or substellar, an unlikely
explanation for the group as a whole.

\subsection{Evidence for `inside out' disk evolution \label{evolution}}

The transition from the class II to the class III stage has been the
focus of a number of recent theoretical investigations. Different
physical processes that could be responsible for disk evolution will
produce different signatures in the color plane as a function of
time. There are 3 classes of models that we will investigate here.
The first is the homologous disk evolution of \citet{wood02}, in which
the disk structure remains constant and the only property that evolves
is the disk mass. The second is the magnetospheric disk accretion
models, which investigate the viscous evolution of disks in the
presence of magnetospheric interactions between a disk and a star
\citep[e.g.,][]{kyh96, armitage99}. Finally, there are models in which
the disk evolution is driven by the competing effects of the ionizing
flux from the central star, which creates a disk wind, and by the
viscosity \citep[][]{hollenbach94, clarke01}. In this section we will
use the observed properties of the passive disk systems to constrain
which model (and associated physical processes) is most likely
responsible for disk evolution.  The observed passive disk properties
we will use are (a) the evidence for an inner
disk hole, as determined through photospheric \kl colors and optically
thick, disk-like \kn colors, (b) the fact that these systems only occur
around M type stars, and (c) the observed fraction of such systems:
10\% of the entire class II sample (including both singles and
binaries), or, more accurately, 0\% for K type class II stars and 15\%
for M type class II stars.

\citet{wood02} model the near- and mid-infrared colors of a typical T
Tauri disk under the assumption that the disk evolves through a
reduction in the disk mass while maintaining a constant disk
structure. They find that the observed spread of \kl and \kn colors
can, in general, be reproduced by disks of different masses observed
at varying disk inclinations. However, Wood et al. point out that the
constant disk structure in this model assumes that disks with inner
holes will not develop as the disks evolve. With $\sim$10\% of our
class II stars showing evidence for an inner disk hole, we can reject
this model.

Models that investigate the viscous evolution of magnetospheric T
Tauri disks \citep[e.g.,][]{kyh96, armitage99} find that disks evolve
in an inside-out manner. The radius of the inner edge of a
magnetically truncated disk is determined by the balance between the
magnetic torques and viscous torques in the disk. As the system
ages and the mass accretion rate drops, the radius at which this
balance occurs increases, hence, an inner disk hole forms and evolves
with the mass accretion rate and magnetic field strength. The disks
are predicted to evolve in the \kl vs \kn color plane through a phase
with photospheric \kl and disk-like \kn colors, suggesting that the
passive disk systems could represent a population that is in the
process of dispersing their inner disk
material. \citet[][]{hartmann98} shows that the size of the inner disk
hole is dependent on a number of parameters (such as stellar radius,
accretion rate and magnetic field strength), most of which are
themselves dependent on the stellar mass. Because of the complexity in
the interplay between the parameters, the exact dependence remains
unknown, however, and so the presence of a spectral type dependence in
the observations, while not inconsistent with this model, cannot
currently constrain it.  The model does fail, however, to accurately
predict the observed fraction of such systems in the sample as a
whole. In this model, disk evolution is driven by the decline in
accretion rate over time and occurs on the viscous timescale of the
entire disk (which scales with the outer radius of the disk and
viscosity parameter as $R_{outer}^{2}/\nu$).Calculations of the
observed disk colors as a function of time predict that roughly 50\%
of all class II stars should be detected with photospheric \kl colors
and \kn $>$ 2, a factor of 5 too large compared to our observations.

\citet{clarke01} investigate disk evolution with the inclusion of disk
photoevaporation. An ionizing flux from the central star creates a
disk wind that is launched beyond some critical radius, $R_{g}= GM_{*}
/ c_{s}^{2}$.  As the mass accretion rate declines with time, there
comes a point when the mass loss in the disk wind at $R_{g}$ exceeds
the mass accretion into the disk region inside that radius. The lack
of replenishment in the inner disk region causes it to evolve on its
own short viscous timescale, producing a rapid removal of material
\citep[][]{clarke01, acp03}. This provides the potential for a rapid
transition in disk evolution, with the inner disk region clearing out
on a timescale on the order of $10^{5}$ yrs for a 1$M_\odot$ T Tauri
star. The removal of material at radii smaller than $R_{g}$ after the
mass accretion rate has declined yields photospheric colors at
wavelengths shorter than some critical value and optically thick
colors at longer wavelengths, with the critical wavelength depending
only on the size of the inner disk hole. In Clarke et al. (2001) the
photoionizing photons from the central star were assumed to have a
constant flux over time. They state that the source of ionizing
photons cannot be predominantly powered by the accretion flow because
as the accretion rate drops over time, this process will essentially
become ineffective.  Indeed, \citet{matsuyama03} show that if the
source of ionizing flux is from the accretion shock, with the flux
declining over time, a disk gap around $R_{g}$ forms but the inner
disk does not clear out any faster than the outer disk. More recently,
\citet{alexander04a, alexander04b, alexander05} have explored
alternative, constant, sources of ultraviolet (UV) photons and have
concluded that chromospheric far-UV photons in T Tauri stars
potentially can provide a high enough flux, causing the inner disk to
rapidly clear once the accretion rate drops.  Within this scenario,
our sample of passive disks can be interpreted as objects that have
already undergone inner disk clearing. The observed fraction of
passive disks around M type stars (15\%) suggests that this phase of
disk evolution, with a cleared inner disk region, lasts for several
times $10^{5}$ yrs. This model also predicts a strong dependence on
the mass of the central star as the disk radius at which a disk wind
can be launched scales linearly with the stellar mass. The
\citet{clarke01} model predicts that while a 1.5 $M_{\odot}$ star will
have an inner disk hole out to 10 AU, a 0.25 $M_{\odot}$ star will
have an inner disk hole that extends out to only 1--2 AU.
Consequently, the reason we detect passive disks solely around M type
stars is purely a selection effect: because our survey covers only the
2--10 \,\micron\, region, we can only detect disk holes up to $\sim$3
AU in size. We would detect a 1.5 $M_{\odot}$ star with a larger disk
hole as a class III, disk-less, object.  In this framework, we
therefore predict that Spitzer observations across the 3 -- 160 micron
region will identify a number of systems with 5--15 AU inner disk
holes around G and K T Tauri stars, and that the size of the inner
hole will scale linearly with the stellar mass. Furthermore, a direct
consequence of the linear relation between stellar mass and inner hole
size is that the timescale required for the inner disk region to clear
is also dependent on the stellar mass. This is because the inner disk
clears on its own viscous timescale, which scales as $R_{g}^{2}/\nu$.
We therefore predict a higher likelihood of detecting objects in the
process of clearing their inner disk around higher mass T Tauri stars,
a prediction that can be tested with Spitzer.

To summarize, the homologous disk evolution of \citet{wood02} can be
discounted by the detection of the passive disk systems with inner
disk holes and models of magnetospheric disk evolution that evolve on
a purely viscous timescale cannot recreate the small fraction of
observed passive disks. However, models that incorporate the effects
of both viscous evolution and disk photoevaporation can recreate all
the observed properties of these systems.

While the disk photoevaporation evolution models seem to be successful
at describing the evolution of the inner edge of the disk, the
evolution of disk material further out is still not well understood. Beyond the
disk wind launch radius, disk material remains and continues evolving
on a viscous timescale \citep{clarke01}.  In star forming regions
where external photoionization does not cause the disk to evolve in an
outside-in manner, the models predict a remaining outer disk annulus
of gas and dust. However, millimeter surveys of T Tauri stars find
that only \app10\% of all WTTS have associated millimeter continuum
emission \citep[e.g.,][]{duvert00, ob95}. The disk evolution timescale
therefore does not seem to have a strong radial function. Recent work
by \citet{takeuchi05}, suggests that the solution may be the
differential radial motion of dust grains and gas particles in
circumstellar disks; for T Tauri disks with 1\,mm grain sizes, the
dust particles should migrate inward on a timescale an order of
magnitude faster than the gas; by the time the disk has started the
inside-out evolution process, the outer disk is already depleted of
dust grains, leaving a gas-rich disk behind. It is clear that
investigating the outer disk structure of candidate passive disks,
both in dust and gas emission, will help explore this idea
further. Our 18 \,\micron\, data does not have the sensitivity to
investigate this question; of the 4 candidate passive disks observed
at 18\,\micron\, in our sample, only 1 system was wide enough and
bright enough to provide resolved photometry for each component. For
FX Tau B, the passive dust disk does seem to extend out to further
radii.  Future deep mid-infrared images of these systems will be able
to pinpoint the radial distribution of small dust grains.

\section{Summary}

Mid-infrared and near-infrared observations of a sample of 65 binary
systems are presented, increasing the number of T Tauri binary systems
with resolved 10\,\micron\, data by a factor of 5.  We have discovered
a sample of class II stars, with clear evidence for optically thick
disk emission in the mid-infrared, that do not appear to be currently
accreting. The majority of these passive disks have an inner disk
hole, determined through the lack of near-infrared excess, and all of
them occur only around the lowest mass stars (M type) in our sample.
These observations are consistent with models that include inside-out
disk evolution. In particular, the observed frequency and spectral
type/mass dependence of the passive disk systems, in the
2--10\,\micron\, region covered in this survey, are supportive of
models that include a photoevaporative disk wind caused by an ionizing
flux from the central star.  A strong prediction of the
photoevaporation models is that there should be a sample of systems
with larger inner disk holes (5-10 AU) around higher mass stars, with
the size of the disk hole scaling with stellar mass.  Such systems
would not be detectable in the 2--10\,\micron\, study presented here;
a multi-wavelength mid-infrared study of T Tauri stars, which will be
available through Spitzer, will be able to determine whether
photoevaporative disk wind models can adequately describe T Tauri disk
evolution over a range of stellar masses.

\vspace*{0.8in} A portion of this research was done while the author
held a National Research Council Research Associateship Award at the
Jet Propulsion Laboratory. We are grateful to Cathie Clarke for
helpful comments, to Robert Pi\~{n}a for his assistance with OSCIR
observations, and to the referee for a careful reading of this
work. Some of the data presented herein were obtained at the W.M. Keck
Observatory, which is operated as a scientific partnership among the
California Institute of Technology, the University of California and
the National Aeronautics and Space Administration. The Observatory was
made possible by the generous financial support of the W.M. Keck
Foundation.  Support for this work was provided by the NASA
Astrobiology Institute and the Packard Foundation. This publication
makes use of data products from the Two Micron All Sky Survey, which
is a joint project of the University of Massachusetts and the Infrared
Processing and Analysis Center/California Institute of Technology,
funded by the National Aeronautics and Space Administration and the
National Science Foundation. The authors also wish to recognize and
acknowledge the significant cultural role that the summit of Mauna Kea
has always had within the indigenous Hawaiian community.  We are
fortunate to have the opportunity to conduct observations from this
mountain.

\appendix
\section{Line of sight extinction \label{reddening}}
Typically, the line of sight extinction to a star is
estimated by measuring the excess color of the star, in comparison to
that expected from the photosphere, in band passes that are
dominated by photospheric emission rather than disk emission.  Such a
method is not open to us; we lack spectral types for each individual
star in approximately 40\% of the binary systems in our sample. Furthermore, a
number of these systems have not been resolved at optical ($R$- or $I$-
band) wavelengths where the photosphere dominates the flux density.
Instead, a secondary method of determining extinction was
applied. \citet{meyer97}, using the sample of T Tauri stars from
\citet{strom89}, found that the classical T Tauri stars (CTTS),
de-reddened using \av determined from the $R-I_{c}$ excesses, defined
a narrow locus in the $(J-H)_{CIT}$ vs. $(H-K)_{CIT}$ color-color
plane. We therefore compared the near-infrared colors of the observed
binary systems with the CTTS locus defined by Meyer et al. (1997) in
order to estimate the line of sight extinction.

Using the 2MASS database, J,H,K photometry was obtained for stars in
the sample.  For most stars, the 2MASS data does not resolve the
binary components and therefore provides composite fluxes only. The
magnitudes obtained from 2MASS were converted into CIT magnitudes
using the filter transformations from \citet{carpenter01}.  Using the
equation in \S3.1 of \citet{pgs03}, which relates the visual
extinction with the distance between the CTTS locus and the stellar
position in the $(J-H)_{CIT}$ vs. $(H-K)_{CIT}$ color-color diagram,
the visual extinction, \av, was calculated for each binary system in
our sample. The \av values calculated using this method are given in
Table \ref{colors}. A few stars did not have 2MASS photometry and some
(18\% of sample) had initial positions that were below the CTTS locus,
possibly the result of scattering from circumstellar material or
photometric variability. As the \av values cannot be determined in
this case, we have arbitrarily set them to have \av = 0.0. The
observed mid-infrared and near-infrared colors are de-reddened using
the \citet{rl} extinction coefficients for the $K$, $L$ and [N] band
fluxes and using the 18\,\micron\, extinction coefficient from Allen's
Astrophysical Quantities.

\section{Absolute Photometric Calibration \label{calib}}

During each night's observations, a number of photometric standards
were observed. While these standards have known flux densities
for a number of mid-infrared filters, this project has used a large
number of both wide and narrow band filters that differ slightly from
those in the literature. We have therefore re-calibrated the standards
in each of the filters used. The standards can be split into two
groups; `primary' standards, that have absolutely calibrated
photometry and a template stellar spectra, available through the
Cohen-Walker radiometric network \citep{cohen92, cohen95, cohen99},
and `secondary' standards that have their absolute calibration
estimated through relative photometry with any primary calibrators
observed on the same night.

Absolute calibration of the primary standards was done by
taking the published, absolutely calibrated, stellar templates and
integrating the stellar spectra ($F_{\lambda}$) through the filter
bandpass ($B_{\lambda}$), the instrument detector QE
($DQE$), and the atmospheric transmission ($T_{atmos}$):

\begin{center}
$F_{filter} = \int \frac{F_{\lambda}*T_{atmos}*DQE*B_{\lambda}}{T_{atmos}*DQE*B_{\lambda}} d\lambda$
\end{center}

DQE and filter transmission curves for the LWS and OSCIR instruments
were provided by R. Campbell and S. Fisher, respectively.  The
model atmosphere used is the ATRANS atmospheric transmission model by
\citet{Lord} and includes 1 mm of precipitable water. These models are
available on the Gemini
website.\footnote{http://www.gemini.edu/sciops/ObsProcess/obsConstraints/ocTransSpectra.html}
Models with precipitable water varying from 1 to 5 mm were tested. For
all bandpasses used in this study, changing the atmospheric water
content changed the total weighted flux by less than 1\%, well below
the observational uncertainties of the data. The resulting flux
densities for these primary photometric calibrators, in Janskys, are
reported in Table \ref{primphot} for each filter used.  Using the
subset of the observed standards that also have fluxes estimated by
the OSCIR team to compare the measured zeropoints, we find that the
flux densities are consistent to within $\sim$6\% and $<$1\%, for the
N band filter and IHW18 filter respectively.

The secondary standards are calibrated through relative photometry to
primary standards that are observed on the same night, under
photometric conditions.  The observed flux
ratios are converted into fluxes using the primary standard
fluxes. The average and standard deviation of the measured fluxes for the 
secondary standards are also
provided in Table \ref{primphot}.  For those standards observed only
during non-photometric nights we have adopted the OSCIR team's
photometry, if available. These are indicated by fluxes placed inside
square brackets.

Throughout this paper, we assume $\alpha$ Lyr has 0.00 mag in each
bandpass used.  While Vega does have a circumstellar shell, the excess
emission from this dust does not occur at wavelengths shorter than
20\,\micron\, \citep[e.g.,][]{allen}.

\section{Known single T Tauri stars \label{app_singles}}

A comparison sample of T Tauri stars which have been determined to be
single has been compiled from Table 7 from \citet{wg01} for Taurus
sources and from \citet{barsony03} for Ophiuchus sources. We only include
sources that have their single status determined using high-resolution
observations, including speckle interferometry (Leinert et al. 1993;
Ghez et al. 1993; Ageorges et al. 1997; Barsony et al. 2003) and lunar
occultation and imaging (Simon et al. 1995). A literature search on
these known single stars was made in order to obtain $K$, $L$ and $N$
band information, as well as accretion information from either \Ha or
\brg line emission. The total sample is comprised of 27 objects that have
near- and mid-infrared disk color indices, an \Ha measurement of
accretion, and an estimate of the visual extinction from 2MASS
photometry.  These are listed in Table \ref{singles}, along with the
spectral type, estimated visual extinction (see Appendix
\ref{reddening} for method), observed colors, and equivalent width
of \Ha line emission for the star. The reference source for
multiplicity is given in column 9, and the observational method used
listed in column 12.

\section{The Passive Disk Candidates \label{passive_obj}}

In this section we provide a summary of each passive disk candidate
presented in this work, identified through the observed combination of
a class II-like \knobs excess and lack of accretion. As pointed out in
\S\ref{discuss}, our analysis rests on the statistical nature of this
sub-set as a whole, however, for any one of these objects the observed
properties may be caused by one of the other possibilities we have
raised (i.e., varying accretion flow, the presence of a nearby
companion). 

{\it WSB 28} -- This 5\arcsec\, binary system in Ophiuchus is one of
the smallest mass ratio systems \citep[q \app 0.1, according to
the][tracks]{baraffe98, baraffe02} in our sample. It is also the
only non-mixed binary system that contains a passive disk; both
components are WTTS as indicated by a lack of \brg emission
\citep{pgs03}. No \Ha equivalent widths have been measured for either
individual component. \citet{wsb} do, however, list the primary as
having a weak \Ha measurement. K-band photometry of the primary has
been obtained on 3 separate occasions, including this work; the object
appears to be photometrically stable ($\sigma_{K}$ = 0.03 mag;
standard deviation from 2MASS; Prato Greene \& Simon 2003 and this
work). Along with \citet{pgs03}, we have obtained K-band photometry on
the secondary; this component also appears stable within the
observational uncertainties. There are a number of line of sight
extinction measurements, obtained using different methods; the CTTS
locus method applied here finds an \av of 3.3 mag for the system as a
whole, similar to that found by \citet{bontemps01} (\av=4.3 mag). Both
combined measurements are intermediate in value between the \av
determined for each individual component, measured by spectral
fitting, which finds an \av of 5.1 \sig 0.6 mag and 2.5 \sig 1
mag for the primary and secondary, respectively
\citep{pgs03}. De-reddening the observed colors with any of the above
\av measurements do not change the classifications for each object;
the primary continues to have no excess in either \kl or \knobs
whereas the secondary, while lacking a \kl excess, shows \knobs $>$ 2,
indicating that this object may have an inner disk hole. While the
observational uncertainties on the color of the secondary are fairly
large, it would take a 3.4\sigm change for the object to fall in the
class III region and cease to be regarded as a passive disk.

{\it DOAR 26} -- This M4/M6 2\farcs2 binary system is a mixed pair:
the primary is a CTTS (EW(\brg) = 2.2 \AA) and the secondary a WTTS
(non-detection of \brg; Prato, Greene \& Simon 2003). No \Ha
measurements that resolve both components are available; the
\citet{wsb} \Ha survey qualitatively measures an \Ha level of 2 (on a
scale of 'weak' to '5') for the combined binary system.  A comparison
of all resolved ground-based near-infrared photometry on this binary
system finds that the K-band flux of both components is stable, with
an rms of 0.1 mag for each component \citep[][this work]{koresko02,
pgs03}. Resolved L-band magnitudes of this system are provided only by
\citet{koresko02} and this work. \citet{koresko02} finds \kl = 0.7
\sig 0.3 for the secondary, which differs by 2\sigm from the value we
find and is associated with much larger observational uncertainties.
Additional, near-simultaneous, K- and L-band measurements are
desirable to explore how variable this system is.  The \av toward the
system as a whole, measured using the CTTS locus method, is 1.64 mag,
intermediate between the values measured from spectral fits for the
primary and secondary \citep[\av = 3.3 \sig 1.2 mag and 1 \sig 0.9
mag, respectively;][]{pgs03}. The de-reddened \knobs colors of this
CTTS-WTTS system are class II-like for both objects, the class II
classification does not change no matter which \av measurement listed
above is used.  The de-reddened \kl colors clearly indicate an excess
for the primary and none for the secondary. While the primary is
clearly a class II-CTTS object, the secondary shows evidence for an
inner disk hole and a lack of accretion.

{\it FQ Tau} -- This system appears to be highly variable.
\citet{hk03} find that it is an M3/M3.5 0\farcs73 binary, with both
components showing \Ha in emission, with an equivalent width of 110
\sig 5 and 23 \sig 4 \AA\, for the primary and secondary,
respectively. On the basis of \Ha alone, the secondary would be
classified as a borderline CTTS/WTTS. \citet{hk03}, however, also use
the lack of [O I] line emission and veiling in the secondary to
classify FQ Tau B as a WTTS, which is the designation we use
here. Photometrically, this system varies considerably,
with the brightest component at both K- and L-band inverting between
the results presented in \citet{wg01} and those in Table\,\ref{photres}. During
1997, \citet{wg01} found the south-western component to be the
brightest at 2.2\,\micron, with observed (and not de-reddened)
near-infrared colors of \kl = 0.43 \sig 0.11 mag and \kl=0.37 \sig
0.11 mag for the primary and secondary, respectively. In comparison,
during 2001, we found the north-eastern component to be brighter by
\app0.3 mag in the K-band, with observed (not de-reddened)
near-infrared colors of \kl = 0.61 \sig 0.07 mag and \kl = 0.75 \sig
0.06 mag for the primary and secondary, respectively. The \av
measurement using the CTTS locus (\av=0.35 mag) is smaller than those
measured from a fit to the optical spectra (\av = 1.95 and 1.8 mag for
the primary and secondary respectively; Hartigan \& Kenyon 2003). The
de-reddened \knobs colors are class II for both components, as are the
\kl colors; no evidence for an inner disk hole is found using our more
recent K- and L- band photometry. De-reddening with either set of \av
measurements does not change the classification for either
component. We consider this object to be a borderline candidate
passive disk and recommend follow-up observations.

{\it FX Tau} -- \citet{duchene99} resolved the individual components
in this 0\farcs85 binary system spectroscopically, finding spectral
types of M1 and M4 for the primary and secondary, respectively, with
an uncertainty of 2 subclasses for each designation.  They found an
\Ha equivalent width of 13 and 1 \AA\, for the primary and secondary
respectively, indicating that this is a mixed CTTS-WTTS binary system.
The resolved \Ha measurements are consistent with previous \Ha
measurements on the combined binary system \citep[14.5
\AA\,;][]{kenyon98}; Pa $\beta$ and \brg are also seen in emission
\citep{muzerolle98, folha01}.  The resolved K-band photometry used is
from \citet{wg01}, which provides an average of all resolved
magnitudes available in the literature.  As such, the observed \kl
color given in Table\,\ref{photres} was not derived from simultaneous
measurements. Our observed L-band magnitudes, 7.6 \sig 0.05 mag and
8.65 \sig 0.06 mag for primary and secondary, respectively, agree well
with the \citet{wg01} measurements (7.49 \sig 0.1 and 8.67 \sig 0.1 mag,
respectively).  The line of sight extinction, measured using the CTTS
locus method, has an \av of 2.24 mag, similar to the value (\av = 1.08
mag) measured from the optical colors given in \citet{kh95}. The
de-reddened \kl and \knobs colors in Table\,\ref{colors} are clearly
class II colors, suggesting that this object maybe in fact undergoing
variable accretion as discussed in \S\ref{discuss}.
  
{\it LkHa 332} -- The 0\farcs34 binary system was spectroscopically
resolved by \citet{hk03} into a K7/M2.5 pair, with \Ha equivalent
widths of 10.9 \sig 0.6 \AA\, and 6.2 \sig 0.6 \AA\, respectively,
making it a CTTS-WTTS mixed pair, although the CTTS classification for
the primary is borderline. This object appears to undergo significant
variation in \Ha emission; previous narrow band \Ha WFPC2 observations
found equivalent widths of 45 \AA\, and 17 \AA\, for the primary and
secondary, respectively, which, again, would classify it as a
CTTS-WTTS pair \citep{wg01}. However, classification of this object in
the literature is inconsistent. Both \citet{hk03} and \citet{wg01}
classify this binary system as a CTTS-CTTS.  According to Tables 3 and
4 in \citet{hk03}, veiling is not present at a $>$2\sigm level in
either component (note that this is contrary to their Table 7 which
states that it is present for the secondary component -- we assume
this is a typo). The only emission line evidence for active accretion
in the secondary is the presence of the the 6300 \AA\, [0 I] line,
which is strangely absent in the CTTS primary. Further spectroscopic
follow up on this close binary system is needed. Using the \Ha
measurements, we refer to this as a CTTS-WTTS system. For calculating
the near-infrared color, we have combined the 1997 K-band measurement
of this object from \citet{woitas01b} with the L-band average
measurement provided by \citet{wg01}; we do this rather than use an
average K-band measurment to minimize any photometric
variability. This provides an observed \kl color of 1.06 \sig 0.1 mag
and 0.03 \sig 0.14 mag for the primary and secondary, respectively.
The average observed \kl colors from \citet{woitas01b} for these
objects (0.97 \sig 0.11 mag and 0.32 \sig 0.31 mag respectively) shows
that photometric variability is likely present; near-simultaneous K-
and L-band photometry on this system would clearly be useful. The
line-of sight extinction estimated through the CTTS locus method is
\av = 3.58 mag, which is approximately 1 mag larger than the \av found
through a fit to the spectroscopic data by Hartigan \& Kenyon (\av =
2.8 mag and 2.3 mag for the primary and secondary, respectively). The
de-reddened colors find clear \knobs excesses in both components,
independent of the \av measurement used. The \kl color of the
secondary however shows evidence for an inner disk hole, although the
lack of simultaneous measurements places some uncertainty on this.

{\it FV Tau /c} -- A 0\farcs67 binary system, FV Tau/c has been
spectroscopically resolved into an M2.5/M3.5 pair by \citet{hk03}. The
system appears to be a mixed WTTS-class I pair, which is quite
unusual, with the WTTS designation for the primary coming from \Ha
equivalent width measurements and the lack of veiling and [O I]
emission \citep{wg01, hk03}. The secondary has very large and variable
\Ha equivalent widths \citep{wg01, hk03} and \citet{wg01} classify it
as a high accretion star.  We combine the 1995 K-band measurement from
\citet{woitas01b} with the average L-band photometry from \citet{wg01}
in order to minimize the uncertainty from photometric variability;
these colors do not represent near-simultaneous measurements.  The \av
estimated from the CTTS locus method is 5.23 mag, a value in between
the measured \av from a fit to the optical spectra by \citet{hk03},
who find \av = 3.25 and 7 mag for the primary and secondary,
respectively.  The de-reddened \knobs colors are class II and class I
for the primary and secondary, respectively, independent of the \av
measurement used, whereas the \kl colors of the primary are below the
\kl photospheric cutoff used, suggesting the presence of an inner disk
hole.

\clearpage 


\clearpage

\begin{deluxetable}{lcllc}
\tabletypesize{\footnotesize}
\tablecaption{T Tauri Binary Sample\label{sample}}
\tablewidth{0pt}
\tablehead{
\multicolumn{1}{l}{Object} &\multicolumn{1}{c}{Separation\tablenotemark{a}} &\multicolumn{2}{c}{J2000 coords} &\multicolumn{1}{c}{Ref\tablenotemark{b}}\\
\colhead{} &\colhead{(\arcsec)} &\colhead{RA} &\colhead{Dec} &\colhead{}}
\startdata
{\em Taurus:} & & & & \\
IS Tau      &0.22 &04 33 36.8 &+26 09 50 &2,3,4 \\    
GG Tau A    &0.25  &04 32 30.3 &+17 31 41 &2,3,6 \\   
FS Tau      &0.27  &04 22 02.1 &+26 57 32 &3,4 \\   
IW Tau      &0.27  &04 41 04.7 &+24 51 06 &3,4 \\   
V410 Tau    &0.29  &04 18 31.1 &+28 27 16 &6 \\   
V927 Tau    &0.30  &04 31 23.8 &+24 10 54 &4 \\   
Haro 6-37   &0.30/2.7    &04 46 59.0 &+17 02 38 &1,3 \\   
XZ Tau      &0.30  &04 31 40.0 &+18 13 58 &2,3 \\   
GH Tau      &0.31  &04 33 06.1 &+24 09 45 &2,3 \\   
GN Tau      &0.31  &04 39 20.8 &+25 45 03 &4,5,6 \\   
CZ Tau      &0.33  &04 18 31.6 &+28 16 59 &3 \\   
Lk Ha 332   &0.33  &04 42 07.7 &+25 23 13 &3 \\   
UZ Tau      &0.34/3.78  &04 32 42.9 &+25 52 32 &2,3,4,6 \\   
V807 Tau    &0.41  &04 33 06.7 &+24 09 56 &2,3,4 \\   
LkCa 3      &0.49  &04 14 48.0 &+27 52 35 &2,3 \\   
DD Tau      &0.56  &04 18 31.1 &+28 16 30 &2,3 \\   
HBC 351     &0.61  &03 52 02.3 &+24 39 51 &3 \\   
Haro 6-28   &0.66   &04 35 55.9 &+22 54 36 &3,4 \\ 
VY Tau      &0.66   &04 39 17.4 &+22 47 54 &3,4 \\ 
FV Tau      &0.72   &04 26 53.6 &+26 06 55 &1,2,3,4 \\  
T Tau       &0.73   &04 21 59.4 &+19 32 06 &2,3 \\ 
FV Tau /c   &0.74   &04 26 54.4 &+26 06 52 &3,4 \\ 
FQ Tau      &0.79   &04 19 12.8 &+28 29 36 &3 \\ 
UY Aur      &0.88   &04 51 47.3 &+30 47 14 &2,3 \\ 
FX Tau      &0.90   &04 30 29.6 &+24 26 46 &2,3,4 \\ 
LkCa 7      &1.05   &04 19 41.3 &+27 49 39 &3 \\ 
RW Aur      &1.42    &05 07 49.6 &+30 24 05 &1,2,3,6 \\
GG Tau  B   &1.48    &04 32 30.3 &+17 31 31 &3,6 \\ 
HBC 358/9   &1.58/20.  &04 03 49.2 &+26 10 54 &3 \\  
HBC 356/7   &2.0   &04 03 14.1 &+25 53 00 &8 \\
HBC 411     &2.04   &04 35 41.2 &+24 11 08 &1,3,4 \\  
HK Tau      &2.40   &04 31 50.5 &+24 24 19 &1,3,4 \\  
IT Tau      &2.48   &04 33 54.7 &+26 13 30 &1,4 \\  
DK Tau      &2.53  &04 30 44.3 &+26 01 24 &1,3,4 \\    
UX Tau      &2.7/5.9   &04 30 03.9 &+18 13 49 &1,3 \\  
HN Tau      &3.1   &04 33 39.3 &+17 51 53 &1,3 \\   
V710 Tau    &3.24    &04 31 57.6 &+18 21 37 &1,3 \\   
J4872       &3.5    &04 25 17.6 &+26 17 51 &1 \\  
HV Tau      &4.00    &04 38 35.3 &+26 10 39 &1,3,4,5 \\  
HBC 354/5   &5.0    &03 54 35.6 &+25 37 12 &8 \\  
\hline
{\em Ophiuchus:} & & & & \\
SR 24       &0.20/6.0 &16 26 58.8 &-24 45 37 &1,4,7 \\
NTTS162218-2420   &0.24 &16 25 19.2 &-24 26 52 &2 \\ 
NTTS155913-2233 &0.29 &16 02 10.5 &-22 41 28 &2 \\
SR 12       &0.30  &16 27 19.5 &-24 41 40 &4,7 \\ 
V853 Oph &0.40 &16 28 45.3 &-24 28 17 &2,4 \\ 
ROXs 31     &0.48  &16 27 52.0 &-24 40 48 &4 \\ 
WSB 11      &0.50   &16 21 57.3 &-22 38 16 &1 \\ 
SR 9        &0.59  &16 27 40.3  &-24 22 03 &2 \\ 
WSB 3       &0.60  &16 18 49.5 &-26 32 53 &1 \\  
Haro 1-4    &0.72 &16 25 10.5 &-23 19 14 &1,2 \\  
DoAr 51     &0.77  &16 32 11.8 &-24 40 19 &7 \\ 
NTTS155203-2338\tablenotemark{c}   &0.80 &16 32 11.8 &-24 40 19 &2 \\   
WSB 18      &1.1    &16 24 59.7 &-24 56 01 &1 \\
WSB 26      &1.2 &16 26 18.3 &-25 20 55 &1 \\   
ROX 15      &1.2 &16 26 42.7 &-24 20 29 &1 \\   
AS 205      &1.32 &16 11 31.4 &-18 38 26 &1,2 \\  
WSB 19      &1.5 &16 25 01.9 &-24 59 33 &1 \\   
DoAr 24E    &2.03 &16 26 23.3 &-24 20 58 &1,2,4 \\
DoAr 26     &2.3  &16 26 34.8 &-23 45 41 &1 \\  
WSB 4       &2.8  &16 18 50.3 &-26 10 08  &1 \\  
ROX 43      &4.8  &16 31 20.0 &-24 40 02 &1,4 \\  
WSB 28      &5.1  &16 26 20.9 &-24 08 51 &1 \\  
SR 21 	    &6.7 &16 27 10.2  &-24 19 16 &1,4 \\
\hline
{\em Corona Australis:} & & && \\
S CrA       &1.3  &19 01 08.5  &-36 57 20 &1 \\   
HBC 679     &4.5  &19 02 22.4 &-36 55 41 &1 \\  
\enddata
\\
\tablenotetext{a}{Two numbers, 
separated by a slash, highlight a triple system.}
\tablenotetext{b}{Source: [1] - \citet{rz93}
[2] - \citet{ghez93}; 
[3] - \citet{leinert93};
[4] - \citet{simon95}; 
[5] - \citet{simon96};
[6] - \citet{ghez97b};
[7] - \citet{ghez94}; 
[8] - \citet{walter88}}
\end{deluxetable}

\clearpage

\begin{deluxetable}{llll}
\tablewidth{0pt}
\tablecaption{Observation Log\label{obs}}
\tablehead{
\multicolumn{1}{l}{Date (UT)} &
\multicolumn{1}{l}{Telescope} &
\multicolumn{1}{l}{Instrument} &
\multicolumn{1}{l}{Filters} 
}
\startdata
\em{Mid-Infrared} \\
1998 May 9-11 &Keck II &\osc &N, IHW18 \\
1998 May 14-15 &Keck II &\osc &N, IHW18 \\
1999 May 5-6 &Keck II &\osc &N, IHW18 \\
1999 Nov 17-20 &Keck II &\osc &N, IHW18 \\
2001 May 10-12 &Keck I &\lw &8.9, 12.5, 18.7 \\
2002 Nov 12-13 &Keck I &\lw &SiC \\
\em{Near-Infrared} \\
2001 Dec 1 &IRTF &\nsf &K,L \\
2001 May 26-27 &IRTF &\nsf &K,L \\
\enddata
\end{deluxetable}

\clearpage

\begin{landscape}
\begin{deluxetable}{llllllccc}
\setlength{\tabcolsep}{0.02in}
\tabletypesize{\scriptsize}
\tablewidth{0pt}
\tablecaption{Observational Results \label{photres}}
\tablehead{
\multicolumn{1}{l}{} &\multicolumn{1}{c}{Date} &\multicolumn{1}{c}{Filter} 
&\multicolumn{1}{c}{Sepn} &\multicolumn{1}{c}{PA} &\multicolumn{1}{c}{Flux}
&\multicolumn{3}{c}{Magnitude\tablenotemark{a}}  \\
\cline{7-9}
\colhead{Object} &
\colhead{(UT)} &
\colhead{} &
\colhead{(\arcsec)} &
\colhead{($^{\circ}$)} & 
\colhead{Ratio} &
\colhead{Prim.} &
\colhead{Sec.} &
\colhead{Total\tablenotemark{b}}
}
\startdata
{\em Taurus:} & & & & & & & & \\
IS Tau     &2002 Nov 12 &SiC &\nodata &\nodata &\nodata &\nodata &\nodata &5.02 \sig 0.11 \\ 
GG Tau A   &1999 Nov 17 &N &0.24 \sig 0.01 &347 \sig 3 &1.03 \sig 0.04 &5.03 \sig 0.17 &5.07 \sig 0.13 &4.30 \sig 0.15 \\
	   &1999 Nov 17 &IHW18 &\nodata &\nodata &\nodata &\nodata &\nodata &2.67 \sig 0.12 \\ 
FS Tau     &1999 Nov 17 &N &\nodata &\nodata &\nodata &\nodata &\nodata &4.19 \sig 0.18 \\ 
IW Tau     &2001 Dec 01 &L &0.30 \sig 0.01 &185 \sig 1 &1.67 \sig 0.03 &8.62 \sig 0.02 &9.18 \sig 0.02 &8.11 \sig 0.02 \\  
           &2002 Nov 13 &SiC &\nodata &\nodata &\nodata &\nodata &\nodata &8.10 \sig 0.14 \\ 
V410 Tau AC &2002 Nov 13 &SiC &\nodata &\nodata &\nodata &\nodata &\nodata &7.17 \sig 0.13 \\ 
V927 Tau   &2001 Dec 01 &K &0.26 \sig 0.01 &292 \sig 1 &1.64 \sig 0.06 &9.23 \sig 0.02 &9.76 \sig 0.03 &8.71 \sig 0.02 \\
	   &2001 Dec 01 &L &0.29 \sig 0.02 &291 \sig 2 &2.37 \sig 0.2 &7.82 \sig 0.13 &8.76 \sig 0.16 &7.44 \sig 0.15 \\ 
           &2002 Nov 12 &SiC &\nodata &\nodata &\nodata &\nodata &\nodata &8.04 \sig 0.15 \\ 
Haro 6-37 Aa-Ab &2002 Nov 13 &SiC &0.30 \sig 0.01 &181 \sig 3 &1.50 \sig 0.14 &5.04 \sig 0.14 &5.47 \sig 0.10 &4.48 \sig 0.13 \\
Haro 6-37 AB &2002 Nov 13 &SiC &2.61 \sig 0.07 &39 \sig 1 &1.19 \sig 0.20 &4.48 \sig 0.13 &4.68 \sig 0.13 &3.82 \sig 0.15 \\
XZ Tau     &1999 Nov 17 &N &\nodata &\nodata &\nodata &\nodata&\nodata &2.17 \sig 0.15 \\ 
GH Tau     &1999 Nov 17 &N &0.30 \sig 0.01 &111 \sig 2 &1.45 \sig 0.05 &5.75 \sig 0.12 &6.15 \sig 0.09 &5.17 \sig 0.11\\
           &1999 Nov 17 &IHW18 &\nodata &\nodata &\nodata &\nodata &\nodata &3.70 \sig 0.11 \\
GN Tau     &2001 Dec 01 &L &0.36 \sig 0.01 &126 \sig 1 &1.53 \sig 0.03 &8.26 \sig 0.07 &7.80 \sig 0.07 &7.25 \sig 0.07 \\
	   &1999 Nov 17 &N &\nodata &\nodata &\nodata &\nodata &\nodata &6.66 \sig 0.17 \\
CZ Tau     &2001 Dec 01 &K &0.33 \sig 0.01 &88 \sig 2 &5.58 \sig 0.41 &9.51 \sig 0.02 &11.38 \sig 0.08 &9.33 \sig 0.03 \\
	   &2001 Dec 01 &L &0.33 \sig 0.01 &92 \sig 2 &1.71 \sig 0.09 &8.90 \sig 0.08 &9.49 \sig 0.09 &8.40 \sig 0.09 \\
           &1999 Nov 16 &N	&\nodata &\nodata &\nodata &\nodata &\nodata &3.79 \sig 0.13 \\
	   &1999 Nov 17 &IHW18 &\nodata &\nodata &\nodata &\nodata &\nodata &6.06 \sig 0.18 \\
Lk Ha 332  &1999 Nov 17 &N &0.34 \sig 0.01 &210 \sig 1. &1.76 \sig 0.08 &4.96 \sig 0.14 &5.57 \sig 0.12 &4.47 \sig 0.14 \\
	   &1999 Nov 17 &IHW18 &\nodata &\nodata &\nodata &\nodata &\nodata &3.87 \sig 0.12 \\
UZ Tau W    &2001 Dec 01 &L &0.35 \sig 0.01 &4 \sig 1 &1.65 \sig 0.06 &7.68 \sig 0.58 &8.23 \sig 0.59 &7.16 \sig 0.84 \\
            &1999 Nov 16 &N &0.37 \sig 0.01 &3.7 \sig 0.6 &1.17 \sig 0.07 &6.43 \sig 0.08 &6.60 \sig 0.08 &5.76 \sig 0.08 \\
	    &1999 Nov 17 &IHW18 &0.37 \sig 0.01 &2.1 \sig 0.9 &1.33 \sig 0.05 &4.64 \sig 0.10 &4.94 \sig 0.08 &4.03 \sig 0.1\\
UZ Tau EW   &2001 Dec 1 &K &3.5 \sig 0.1 &275 \sig 1 &1.81 \sig 0.08 &7.60 \sig 0.1 &8.24 \sig 0.05 &7.12 \sig 0.09 \\
	    &2001 Dec 1 &L &3.5 \sig 0.1 &276 \sig 1 &2.51 \sig 0.08 &6.78 \sig 0.17 &7.78 \sig 0.13 &6.42 \sig 0.17 \\
            &1999 Nov 16 &N &3.52 \sig 0.06 &272.9 \sig 0.4 &8.02 \sig 0.88 &3.50 \sig 0.09 &5.76 \sig 0.08 &3.37 \sig 0.09 \\
	    &1999 Nov 17 &IHW18 &3.44 \sig 0.06 &273.0 \sig 0.4 &8.79 \sig 1.2 &2.28 \sig 0.11 &4.64 \sig 0.10 &2.16 \sig 0.11 \\
V807 Tau    &1999 Nov 17 &N &\nodata &\nodata &\nodata &\nodata &\nodata &5.27 \sig 0.18* \\
	    &1999 Nov 17 &IHW18 &\nodata &\nodata &\nodata &\nodata &\nodata &4.04 \sig 0.11* \\
LkCa 3      &2001 Dec 01 &K &0.47 \sig 0.01 &72 \sig 1 &1.18 \sig 0.03 &8.06 \sig 0.03 &8.24 \sig 0.03 &7.39 \sig 0.03 \\
	    &2001 Dec 01 &L &0.47 \sig 0.01 &72 \sig 1 &1.17 \sig 0.01 &8.02 \sig 0.03 &8.19 \sig 0.04 &7.35 \sig 0.03 \\
            &1999 Nov 17 &N &\nodata &\nodata &\nodata &\nodata &\nodata &7.52 \sig 0.19\tablenotemark{c} \\
	    &1999 Nov 17 &IHW18 &\nodata &\nodata &\nodata &\nodata &\nodata &\nodata \\
	    &2002 Nov 13 &SiC &0.48 \sig 0.02 &76 \sig 2 &1.41 \sig 0.06 &7.72 \sig 0.11 &8.10 \sig 0.09 &7.14 \sig 0.10\\
DD Tau      &1999 Nov 17 &N &0.54 \sig 0.01 &179 \sig 2 &1.75 \sig 0.08 &4.53 \sig 0.06 &5.13 \sig 0.06 &4.04 \sig 0.06\\
HBC 351     &2001 Dec 01 &K &0.55 \sig 0.01 &314 \sig 1 &3.61 \sig 0.39 &9.20 \sig 0.04 &10.6 \sig 0.1 &8.94 \sig 0.05 \\
	    &2001 Dec 01 &L &0.54 \sig 0.02 &315 \sig 1 &5.14 \sig 0.06 &\nodata &\nodata &\nodata \\
            &2002 Nov 13 &SiC &\nodata &\nodata &\nodata &\nodata &\nodata &\nodata \\
Haro 6-28   &2001 Dec 01 &L &0.65 \sig 0.01 &247 \sig 1 &2.07 \sig 0.08 &8.81 \sig 0.30 &9.60 \sig 0.32 &8.38 \sig 0.36 \\
            &1999 Nov 17 &N &0.64 \sig 0.01 &246.5 \sig 0.4 &0.94 \sig 0.04 &7.38 \sig 0.16 &7.31 \sig 0.18 &6.59 \sig 0.19 \\
            &1999 Nov 17 &IHW18 &0.68 \sig 0.01 &245.1 \sig 0.6 &0.79 \sig 0.07 &5.85 \sig 0.07 &5.58 \sig 0.07 &4.96 \sig 0.07\\
VY Tau      &2001 Dec 01 &L &0.67 \sig 0.01 &318 \sig 1 &6.14 \sig 0.27 &8.47 \sig 0.08 &10.45 \sig 0.11 &8.31 \sig 0.09 \\
            &1999 Nov 17 &N &\nodata &\nodata &\nodata &\nodata &\nodata &6.63 \sig 0.16* \\
FV Tau      &1999 Nov 17 &N &0.71 \sig 0.02 &92.1 \sig 0.8 &2.2 \sig 0.2 &4.07 \sig 0.17 &4.92 \sig 0.12 &3.66 \sig 0.17 \\
	    &1999 Nov 17 &IHW18 &0.67 \sig 0.03 &112 \sig 1 &1.51 \sig 0.06 &3.07 \sig 0.13 &3.52 \sig 0.11 &2.52 \sig 0.13\\
T Tau       &1999 Nov 17 &N &0.67 \sig 0.01 &181.2 \sig 0.4 &0.89 \sig 0.02 &1.54 \sig 0.15 &1.41 \sig 0.16 &0.72 \sig 0.17 \\
            &1999 Nov 17 &IHW18 &0.67 \sig 0.02 &181.6 \sig 0.8 &0.87 \sig 0.01 &-1.48 \sig 0.10 &-1.63 \sig 0.1 &-2.31 \sig 0.1 \\
FV Tau /c   &1999 Nov 17 &N &0.67 \sig 0.02 &112 \sig 1 &0.74 \sig 0.09 &8.59 \sig 0.14 &8.26 \sig 0.17 &7.66 \sig 0.17 \\
	    &1999 Nov 17 &IHW18 &\nodata &\nodata &\nodata &\nodata &\nodata &\nodata \\
	    &2002 Nov 12 &SiC &0.68 \sig 0.02 &115 \sig 1 &0.75 \sig 0.04 &6.27 \sig 0.07 &5.96 \sig 0.10 &5.36 \sig 0.09 \\
FQ Tau      &2001 Dec 01 &K &0.73 \sig 0.01 &67 \sig 1 &0.77 \sig 0.08 &10.15 \sig 0.06 &9.87 \sig 0.05 &9.25 \sig 0.06 \\
	    &2001 Dec 01 &L &0.73 \sig 0.01 &67 \sig 1 &0.68 \sig 0.04 &9.54 \sig 0.04 &9.12 \sig 0.03 &8.56 \sig 0.03 \\
            &1999 Nov 20 &N &0.66 \sig 0.01 &70 \sig 1 &1.13 \sig 0.04 &7.42 \sig 0.04 &7.55 \sig 0.09 &6.73 \sig 0.07 \\
	    &2002 Nov 12 &SiC &0.74 \sig 0.02 &66 \sig 1 &1.05 \sig 0.02 &7.18 \sig 0.09 &7.24 \sig 0.10 &6.46 \sig 0.10 \\
UY Aur      &1999 Nov 16 &N &0.88 \sig 0.01 &227.6 \sig 0.4 &2.07 \sig 0.03 &3.37 \sig 0.1 &4.16 \sig 0.08 &2.94 \sig 0.10 \\
            &1999 Nov 16 &IHW18 &0.87 \sig 0.01 &227.7 \sig 0.4 &2.41 \sig 0.02 &1.28 \sig 0.05 &2.23 \sig 0.04 &0.90 \sig 0.05 \\
FX Tau      &2001 Dec 01 &L &0.85 \sig 0.02 &289 \sig 1 &2.63 \sig 0.08 &7.60 \sig 0.05 &8.65 \sig 0.06 &7.25 \sig 0.05 \\
            &1999 Nov 18 &N &0.87 \sig 0.02 &289.4 \sig 0.4 &2.75 \sig 0.11 &5.36 \sig 0.11 &6.46 \sig 0.06 &5.03 \sig 0.10 \\
	    &1999 Nov 18 &IHW18 &0.84 \sig 0.02 &290.5 \sig 0.5 &2.79 \sig 0.33 &4.01 \sig 0.10 &5.13 \sig 0.03 &3.68 \sig 0.08 \\
LkCa 7      &2001 Dec 01 &K &0.98 \sig 0.02 &25 \sig 1 &1.851 \sig 0.05 &8.66 \sig 0.03 &9.33 \sig 0.04 &8.19 \sig 0.03 \\
	    &2001 Dec 01 &L &0.99 \sig 0.02 &25 \sig 1 &1.866 \sig 0.02 &8.62 \sig 0.01 &9.29 \sig 0.02 &8.15 \sig 0.01 \\
            &1999 Nov 18 &N &\nodata &\nodata &\nodata &\nodata &\nodata &8.23 \sig 0.12* \\
RW Aur      &1999 Nov 16 &N &1.42 \sig 0.02 &254.6 \sig 0.4 &13.63 \sig 0.06 &3.12 \sig 0.09 &5.96 \sig 0.09 &3.04 \sig 0.09\\
            &1999 Nov 16 &IHW18 &1.38 \sig 0.02 &254.7 \sig 0.4 &10.89 \sig 0.46 &1.62 \sig 0.05 &4.22 \sig 0.06 &1.53 \sig 0.05\\
GG Tau  B   &1999 Nov 18 &N &1.43 \sig 0.02 &133.2 \sig 0.4 &6.75 \sig 0.25 &6.72 \sig 0.11 &8.80 \sig 0.11 &6.57 \sig 0.12\\
HBC 358 AB  &2001 Dec 1 &K &1.55 \sig 0.06 &226 \sig 1 &1.777 \sig 0.003 &9.80 \sig 0.06 &10.43 \sig 0.06 &9.32 \sig 0.06 \\
 	    &2001 Dec 1 &L &1.55 \sig 0.06 &224 \sig 2 &1.77 \sig 0.16 &9.79 \sig 0.04 &10.41 \sig 0.14 &9.30 \sig 0.08 \\
            &2002 Nov 13 &SiC &\nodata &\nodata &\nodata &\nodata &\nodata &\nodata \\
HBC 358/HBC 359  &2001 Dec 1 &K &20.1 \sig 0.8 &267 \sig 1 &1.31 \sig 0.08 &9.51 \sig 0.03 &9.80 \sig 0.06 &8.89 \sig 0.04 \\
	    &2001 Dec 1 &L &20.1 \sig 0.8 &267 \sig 1 &1.18 \sig 0.35 &9.61 \sig 0.32 &9.79 \sig 0.04 &8.94 \sig 0.21 \\
HBC 356/357 &2001 Dec 1 &K &1.26 \sig 0.05 &5 \sig 2 &1.04 \sig 0.01 &10.72 \sig 0.03 &10.77 \sig 0.03 &9.99 \sig 0.03 \\
	    &2001 Dec 1 &L &1.25 \sig 0.05 &4 \sig 2 &1.1 \sig 0.1 &10.86 \sig 0.01 &10.92 \sig 0.1 &10.14 \sig 0.06 \\
            &2002 Nov 13 &SiC &\nodata &\nodata &\nodata &\nodata &\nodata &\nodata \\
HBC 411     &2001 Dec 1 &L &2.08 \sig 0.08 &175 \sig 1 &5.99 \sig 0.24 &7.86 \sig 0.01 &9.80 \sig 0.04 &7.69 \sig 0.01 \\
            &2002 Nov 13 &SiC &2.02 \sig 0.05 &175 \sig 1 &10.22 \sig 0.08 &4.68 \sig 0.12 &7.21 \sig 0.12 &4.58 \sig 0.13 \\
HK Tau      &2001 Dec 1 &L &2.39 \sig 0.1 &168.1 \sig 2.7 &25.5 \sig 2.4 &8.08 \sig 0.1 &11.58 \sig 0.16 &8.04 \sig 0.1\tablenotemark{d} \\
            &2002 Nov 13 &SiC &2.23 \sig 0.06 &170 \sig 1 &30 \sig 4 &5.33 \sig 0.07 &9.03 \sig 0.16 &5.30 \sig 0.07 \\
IT Tau      &2001 Dec 1 &L &2.4 \sig 0.1 &225 \sig 2 &3.94 \sig 0.06 &7.67 \sig 0.07 &9.15 \sig 0.05 &7.42 \sig 0.07 \\
            &2002 Nov 13 &SiC &2.31 \sig 0.06 &224 \sig 1 &2.95 \sig 0.01  &5.31 \sig 0.11 &6.48 \sig 0.11 &4.99 \sig 0.11 \\
DK Tau      &2002 Nov 13 &SiC &2.27 \sig 0.06 &117 \sig 1 &8.53 \sig 0.04 &3.42 \sig 0.14 &5.75 \sig 0.13 &3.30 \sig 0.15 \\
UX Tau AB   &2002 Nov 12 &SiC &5.6 \sig 0.1 &268 \sig 1 &11.9 \sig 1.8 &5.62 \sig 0.11 &8.31 \sig 0.12 &5.53 \sig 0.12 \\
UX Tau AC   &2002 Nov 12 &SiC &2.50 \sig 0.3 &181 \sig 5 &59.7 \sig 7.2 &5.62 \sig 0.11 &10.06 \sig 0.07 &5.60 \sig 0.12 \\
HN Tau      &2002 Nov 13 &SiC &\nodata &\nodata &\nodata &\nodata &\nodata &3.53 \sig 0.13* \\
V710 Tau    &2002 Nov 12 &SiC &3.06 \sig 0.08 &178 \sig 1 &12.73 \sig 0.11 &5.44 \sig 0.10 &8.21 \sig 0.10 &5.36 \sig 0.1\\
J4872       &2001 Dec 1 &K &3.4 \sig 0.1 &234 \sig 1 &1.78 \sig 0.06 &8.58 \sig 0.03 &9.20 \sig 0.06 &8.09 \sig 0.04 \\
	    &2001 Dec 1 &L &3.4 \sig 0.1 &234 \sig 1 &1.75 \sig 0.02 &8.53 \sig 0.07 &9.14 \sig 0.1 &8.04 \sig 0.08 \\
	    &2002 Nov 12 &SiC &\nodata &\nodata &\nodata &\nodata &\nodata &\nodata \\
HV Tau      &2001 Dec 1 &K &4.0 \sig 0.2 &43 \sig 1 &73.3 \sig 8.3 &7.86 \sig 0.04 &12.2 \sig 0.1 &7.84 \sig 0.04 \\
	    &2001 Dec 1 &L &3.9 \sig 0.2 &43 \sig 1 &36.4 \sig 1.8 &7.66 \sig 0.04 &11.34 \sig 0.04 &7.62 \sig 0.04 \\
            &2002 Nov 13 &SiC &4.1 \sig 0.1 &43 \sig 1 &1.20 \sig 0.01 &7.59 \sig 0.13 &7.78 \sig 0.14 &6.93 \sig 0.14 \\
HBC 355/354 &2001 Dec 1 &K &6.3 \sig 0.3 &298 \sig 1 &2.18 \sig 0.03 &10.15 \sig 0.01 &10.99 \sig 0.01 &9.74 \sig 0.01 \\
	    &2001 Dec 1 &L &6.2 \sig 0.3 &298 \sig 1 &2.22 \sig 0.1 &10.16 \sig 0.07 &11.03 \sig 0.11 &9.76 \sig 0.09 \\  
            &2002 Nov 13 &SiC &\nodata &\nodata &\nodata &\nodata &\nodata &\nodata \\
\hline
{\em Ophiuchus:} & & & & & & & & \\
SR 24N	&2001 May 26 &K &0.112 \sig 0.002 &63 \sig 2 &1.523 \sig 0.03 &8.50 \sig 0.01 &8.96 \sig 0.01 &7.95 \sig 0.01 \\
	&2001 May 26 &L &0.12 \sig 0.01 &59 \sig 6 &1.887 \sig 0.16 &6.80 \sig 0.04 &7.49 \sig 0.07 &6.34 \sig 0.05 \\ 
SR 24 NS &1999 May 04 &N &5.08 \sig 0.08 &347.7 \sig 0.4 &0.95 \sig 0.01 &3.50 \sig 0.12 &3.45 \sig 0.12 &2.72 \sig 0.13 \\
	 &1999 May 04 &IHW18 &5.01 \sig 0.08 &347.5 \sig 0.4 &1.87 \sig 0.05 &1.87 \sig 0.07 &2.55 \sig 0.05 &1.41 \sig 0.06\\
NTTS162218-2420 &2001 May 26 &K &0.19 \sig 0.004 &147 \sig 2 &1.25 \sig 0.12 &8.29 \sig 0.06 &8.53 \sig 0.07 &7.65 \sig 0.07 \\ 
		&2001 May 26 &L &0.201 \sig 0.004 &145 \sig 1 &1.40 \sig 0.01 &8.22 \sig 0.01 &8.58 \sig 0.01 &7.63 \sig 0.01 \\
                &2001 May 10 &8.8 &\nodata &\nodata &\nodata &\nodata &\nodata &7.32 \sig 0.15 \\
	        &2001 May 10 &12.5 &\nodata &\nodata &\nodata &\nodata &\nodata &\nodata \\
NTTS155913-2233 &1999 May 05 &N &\nodata &\nodata &\nodata &\nodata &\nodata &6.94 \sig 0.12* \\
SR 12           &2001 May 26 &K &0.228 \sig 0.004 &82 \sig 1 &1.292 \sig 0.02 &8.90 \sig 0.01 &9.18 \sig 0.01 &8.28 \sig 0.01 \\
		&2001 May 26 &L &0.237 \sig 0.005 &83 \sig 1 &1.50 \sig 0.02 &8.73 \sig 0.06 &9.17 \sig 0.07 &8.18 \sig 0.07 \\     
		&2001 May 11 &8.8 &\nodata &\nodata &\nodata &\nodata &\nodata &\nodata \\
V853 OPH	&2001 May 26 &K &0.34 \sig 0.01 &93 \sig 1 &3.55 \sig 0.16 &8.22 \sig 0.02 &9.60 \sig 0.04 &7.95 \sig 0.02 \\
		&2001 May 26 &L &0.33 \sig 0.01 &94 \sig 1 &3.91 \sig 0.19 &7.64 \sig 0.03 &9.12 \sig 0.06 &7.39 \sig 0.04 \\
		&1999 May 04 &IHW18 &\nodata &\nodata &\nodata &\nodata &\nodata &3.62 \sig 0.05 \\
		&1999 May 04 &N &0.26 \sig 0.04 &101 \sig 4 &3.9 \sig 0.4 &5.94 \sig 0.09 &7.42 \sig 0.06 &5.69 \sig 0.09 \\
ROXs 31 	&2001 May 26 &K &0.38 \sig 0.01 &252 \sig 1 &1.31 \sig 0.04 &8.68 \sig 0.02 &8.97 \sig 0.03 &8.06 \sig 0.02 \\
		&2001 May 26 &L &0.38 \sig 0.01 &252 \sig 1 &1.65 \sig 0.09 &8.31 \sig 0.03 &8.85 \sig 0.04 &7.79 \sig 0.03 \\
        	&1999 May 05 &N &\nodata &\nodata &\nodata &\nodata &\nodata &\nodata \\
WSB 11  	&2001 May 27 &K &0.47 \sig 0.01 &317 \sig 1 &2.22 \sig 0.06 &10.39 \sig 0.01 &11.26 \sig 0.02 &9.99 \sig 0.01 \\
        	&1998 May 10 &N &0.48 \sig 0.01 &314.9 \sig 0.6 &0.94 \sig 0.10 &\nodata &\nodata &\nodata \\
		&1999 May 04 &N &0.48 \sig 0.01 &317 \sig 1 &1.09 \sig 0.18 &7.64 \sig 0.06 &7.74 \sig 0.06 &6.94 \sig 0.06 \\
		&1999 May 04 &IHW18 &\nodata &\nodata &\nodata &\nodata &\nodata &\nodata \\
SR 9             &2001 May 26 &K &0.59 \sig 0.01 &353 \sig 1 &10.82 \sig 0.16 &7.18 \sig 0.02 &9.77 \sig 0.03 &7.08 \sig 0.02 \\
		 &2001 May 26 &L &0.61 \sig 0.02 &353 \sig 1 &13.88 \sig 0.55 &6.56 \sig 0.02 &9.41 \sig 0.06 &6.48 \sig 0.02 \\
                 &1999 May 04 &N &0.59 \sig 0.02 &356.8 \sig 0.7 &9.54 \sig 0.9 &4.37 \sig 0.09 &6.82 \sig 0.11 &4.26 \sig 0.10\\
WSB 3            &2001 May 27 &K &0.59 \sig 0.01 &159 \sig 1 &2.36 \sig 0.01 &9.47 \sig 0.01 &10.40 \sig 0.01 &9.09 \sig 0.01 \\
		 &2001 May 27 &L &0.60 \sig 0.02 &159 \sig 1 &4.42 \sig 0.24 &9.07 \sig 0.04 &10.68 \sig 0.08 &8.85 \sig 0.05 \\
                 &2001 May 10 &8.8 &\nodata &\nodata &\nodata &\nodata &\nodata &\nodata \\
Haro 1-4         &2001 May 26 &K &0.74 \sig 0.01 &30 \sig 1 &3.37 \sig 0.07 &7.58 \sig 0.02 &8.90 \sig 0.03 &7.30 \sig 0.02 \\
		&2001 May 26 &L &0.75 \sig 0.01 &30 \sig 1 &5.88 \sig 0.02 &6.83 \sig 0.02 &8.75 \sig 0.02 &6.66 \sig 0.02 \\
                &2001 May 11 &8.8 &0.73 \sig 0.02 &27 \sig 1 &3.17 \sig 0.5 &\nodata &\nodata &\nodata \\
		&2001 May 11 &12.5 &0.74 \sig 0.02 &26 \sig 1 &3.14 \sig 0.3 &\nodata &\nodata &\nodata \\
		&1998 May 09 &N &0.76 \sig 0.02 &27.2 \sig 0.6 &1.77 \sig 0.06 &\nodata &\nodata &\nodata \\
		&1998 May 09 &IHW18 &0.76 \sig 0.01 &28.8 \sig 0.4 &2.51 \sig 0.03 &\nodata &\nodata &\nodata \\
DoAr 51        &2001 May 27 &K &0.75 \sig 0.02 &79 \sig 1 &9.72 \sig 2.64 &7.75 \sig 0.07 &10.22 \sig 0.27 &7.64 \sig 0.09 \\
	       &2001 May 27 &L &0.72 \sig 0.02 &79 \sig 2 &3.41 \sig 0.41 &7.86 \sig 0.14 &9.19 \sig 0.21 &7.58 \sig 0.17 \\
               &1998 May 10 &N &0.79 \sig 0.02 &77.6 \sig 0.4 &0.51 \sig 0.09 &\nodata &\nodata &\nodata \\
NTTS155203-2338\tablenotemark{e}  &2001 May 26 &K &0.62 \sig 0.03 &231 \sig 1 &3.76 \sig 0.06 &7.21 \sig 0.03 &8.65 \sig 0.05 &6.95 \sig 0.03 \\
		 &2001 May 26 &L &0.73 \sig 0.03 &233 \sig 2 &5.68 \sig 0.5 &6.69 \sig 0.03 &8.57 \sig 0.08 &6.51 \sig 0.04 \\
                 &2001 May 10 &8.8 &0.72 \sig 0.02 &232 \sig 1 &5.34 \sig 0.23 &7.10 \sig 0.05 &8.92 \sig 0.02 &6.92 \sig 0.05 \\
		 &2001 May 10 &12.5 &\nodata &\nodata &\nodata &7.49 \sig 0.11 &\nodata &7.49 \sig 0.11 \\
		 &1998 May 09 &N &0.73 \sig 0.02 &233.5 \sig 0.8 &4.36 \sig 0.6 &\nodata &\nodata &\nodata \\
WSB 18           &2001 May 27 &K &0.97 \sig 0.11 &80 \sig 1 &1.15 \sig 0.3 &10.2 \sig 0.2 &10.38 \sig 0.08 &9.53 \sig 0.16 \\
		 &2001 May 27 &L &1.04 \sig 0.05 &83 \sig 1 &0.841 \sig 0.08 &9.94 \sig 0.13 &9.75 \sig 0.03 &9.09 \sig 0.08 \\
                 &2001 May 12 &8.8 &1.03 \sig 0.03 &85 \sig 1 &3.87 \sig 0.09 &9.21 \sig 0.08 &7.74 \sig 0.1 &7.49 \sig 0.1 \\
        	&2001 May 12 &12.5 &1.01 \sig 0.03 &77 \sig 3 &4.89 \sig 0.23 &8.86 \sig 0.08 &7.14 \sig 0.08 &6.94 \sig 0.08 \\
WSB 26           &2001 May 27 &K &1.09 \sig 0.05 &26 \sig 2 &1.24 \sig 0.01 &9.87 \sig 0.09 &10.10 \sig 0.08 &9.23 \sig 0.09 \\
		 &2001 May 27 &L &1.10 \sig 0.05 &25 \sig 1 &1.31 \sig 0.04 &9.13 \sig 0.03 &9.42 \sig 0.01 &8.51 \sig 0.02 \\
                 &1999 May 04 &N &1.14 \sig 0.02 &24.6 \sig 0.4 &0.41 \sig 0.03 &7.25 \sig 0.11 &6.28 \sig 0.11 &5.91 \sig 0.12 \\
		 &1998 May 11 &N &1.12 \sig 0.02 &23.2 \sig 0.7 &3.78 \sig 0.2 &\nodata &\nodata &\nodata \\
		 &1999 May 04 &IHW18 &\nodata &\nodata &\nodata &\nodata &\nodata &\nodata \\
ROX 15           &2001 May 27 &K &1.40 \sig 0.06 &69 \sig 1 &3.15 \sig 0.05 &8.22 \sig 0.09 &9.47 \sig 0.07 &7.92 \sig 0.09 \\
		 &2001 May 27 &L &1.40 \sig 0.06 &69 \sig 1 &2.90 \sig 0.1 &7.39 \sig 0.08 &8.55 \sig 0.04 &7.07 \sig 0.07 \\
		 &1998 May 11 &N &1.39 \sig 0.02 &67.6 \sig 0.4 &4.86 \sig 0.4 &\nodata &\nodata &\nodata \\
		 &1999 May 05 &N &1.45 \sig 0.02 &68.7 \sig 0.4 &3.97 \sig 0.18 &5.71 \sig 0.05 &7.21 \sig 0.09 &5.47 \sig 0.06\\
                 &1999 May 05 &IHW18 &\nodata &\nodata &\nodata &\nodata &\nodata &\nodata \\
AS 205           &2001 May 27 &K &1.30 \sig 0.07 &214 \sig 1 &2.81 \sig 0.06 &5.95 \sig 0.1 &7.07 \sig 0.13 &5.62 \sig 0.11 \\
		 &2001 May 27 &L &1.32 \sig 0.05 &213 \sig 1 &2.90 \sig 0.23 &4.9 \sig 0.1 &6.05 \sig 0.02 &4.58 \sig 0.08 \\
                 &2001 May 10 &8.8 &1.30 \sig 0.03 &210 \sig 1 &2.74 \sig 0.02 &2.46 \sig 0.15 &3.55 \sig 0.15 &2.12 \sig 0.16 \\
		 &2001 May 10 &12.5 &1.30 \sig 0.03 &210 \sig 1 &3.48 \sig 0.07 &1.47 \sig 0.15 &2.82 \sig 0.14 &1.19 \sig 0.16 \\
		&1998 May 11 &N &1.31 \sig 0.04 &212.1 \sig 0.6 &4.44 \sig 0.5 &\nodata &\nodata &\nodata \\
WSB 19           &2001 May 27 &K &1.48 \sig 0.06 &262 \sig 1 &2.12 \sig 0.1 &9.84 \sig 0.04 &10.66 \sig 0.09 &9.42 \sig 0.06 \\
		 &2001 May 27 &L &1.49 \sig 0.06 &262 \sig 1 &2.68 \sig 0.03 &9.13 \sig 0.02 &10.20 \sig 0.04 &8.79 \sig 0.03 \\
		 &1999 May 05 &N &1.51 \sig 0.02 &263.2 \sig 0.4 &2.38 \sig 0.06 &5.72 \sig 0.06 &6.67 \sig 0.08 &5.34 \sig 0.07 \\
                 &1999 May 05 &IHW18 &\nodata &\nodata &\nodata &3.83 \sig 0.16 &\nodata &3.83 \sig 0.16 \\
DoAr 24E         &2001 May 27 &K &1.98 \sig 0.08 &150 \sig 1 &2.30 \sig 0.5 &7.13 \sig 0.14 &8.0 \sig 0.1 &6.73 \sig 0.14 \\
		 &2001 May 27 &L &1.98 \sig 0.08 &150 \sig 1 &0.78 \sig 0.03 &6.42 \sig 0.09 &6.15 \sig 0.14 &5.52 \sig 0.12 \\
                 &1998 May 10 &N &2.00 \sig 0.03 &149.6 \sig 0.4 &0.35 \sig 0.004 &\nodata &\nodata &\nodata \\
                 &1998 May 10 &IHW18 &2.01 \sig 0.03 &149.9 \sig 0.4 &0.40 \sig 0.004 &\nodata &\nodata &\nodata \\
DoAr 26\tablenotemark{f}    &2001 May 27 &K &2.23 \sig 0.09 &131 \sig 1 &3.60 \sig 0.14 &9.1 \sig 0.3 &10.4 \sig 0.2 &8.81 \sig 0.32 \\
		 &2001 May 27 &L &2.22 \sig 0.09 &132 \sig 2 &6.17 \sig 0.01 &8.28 \sig 0.05 &10.25 \sig 0.06 &8.12 \sig 0.05 \\
                 &1998 May 11 &N &2.27 \sig 0.04 &130.6 \sig 0.5 &8.16 \sig 0.9 &\nodata &\nodata &\nodata \\
		 &1999 May 05 &N &2.29 \sig 0.04 &131.4 \sig 0.4 &5.87 \sig 0.10 &6.54 \sig 0.06 &8.46 \sig 0.06 &6.37 \sig 0.06 \\
		 &1999 May 05 &IHW18 &\nodata &\nodata &\nodata &\nodata &\nodata &\nodata \\
WSB 4            &2001 May 27 &K &2.78 \sig 0.11 &130 \sig 1 &0.611 \sig 0.002 &10.92 \sig 0.12 &10.38 \sig 0.12 &9.86 \sig 0.13 \\
		 &2001 May 27 &L &2.77 \sig 0.12 &130 \sig 1 &0.394 \sig 0.02 &10.65 \sig 0.12 &9.64 \sig 0.13 &9.28 \sig 0.14 \\
                 &2001 May 12 &8.8 &\nodata &\nodata &\nodata &\nodata &\nodata &8.18 \sig 0.10* \\
                 &2001 May 12 &12.5 &\nodata &\nodata &\nodata &\nodata &\nodata &8.18 \sig 0.07* \\
ROX 43     	&2001 May 26 &K &4.41 \sig 0.18 &12 \sig 1 &2.70 \sig 0.03 &6.66 \sig 0.01 &7.74 \sig 0.02 &6.32 \sig 0.01 \\
		&2001 May 26 &L &4.40 \sig 0.18 &12 \sig 1 &4.78 \sig 0.06 &5.51 \sig 0.01 &7.221 \sig 0.001 &5.31 \sig 0.01 \\
		&2001 May 11 &8.8 &\nodata &\nodata &\nodata &\nodata &\nodata &\nodata \\
		&2001 May 11 &12.5 &\nodata &\nodata &\nodata &\nodata &\nodata &\nodata \\
WSB 28     	&2001 May 27 &K &5.00 \sig 0.2 &358 \sig 2 &10.67 \sig 3.36 &9.540 \sig 0.004 &12.08 \sig 0.3 &9.44 \sig 0.03 \\
		&2001 May 27 &L &5.0 \sig 0.2 &358 \sig 1 &9.18 \sig 1.26 &9.49 \sig 0.13 &11.89 \sig 0.3 &9.38 \sig 0.16 \\
                &2001 May 12 &8.8 &4.9 \sig 0.1 &356 \sig 1 &1.24 \sig 0.04 &9.41 \sig 0.11 &9.64 \sig 0.08 &8.77 \sig 0.10 \\
SR 21	  	&1998 May 11 &N &7.67 \sig 0.12 &177.9 \sig 0.4 &10.84 \sig 5.27 &\nodata &\nodata &\nodata \\
	  	&1998 May 11 &IHW18 &7.85 \sig 0.13 &179.1 \sig 0.4 &3.43 \sig 1.2 & \nodata&\nodata &\nodata \\
\hline
{\em Corona Australis:} & & & & & & & & \\
S CRA            &2001 May 26 &K &1.30 \sig 0.05 &150 \sig 1 &3.27 \sig 0.1 &6.04 \sig 0.09 &7.33 \sig 0.13 &5.75 \sig 0.10 \\
		 &2001 May 26 &L &1.30 \sig 0.05 &150 \sig 1 &3.44 \sig 0.1 &4.62 \sig 0.06 &5.97 \sig 0.03 &4.34 \sig 0.05 \\
                 &2001 May 12 &8.8 &1.27 \sig 0.03 &148 \sig 1 &1.43 \sig 0.01 &3.51 \sig 0.11 &3.90 \sig 0.11 &2.93 \sig 0.12 \\
	         &2001 May 12 &12.5 &1.27 \sig 0.03 &148 \sig 1 &1.32 \sig 0.01 &2.88 \sig 0.07 &3.18 \sig 0.07 &2.27 \sig 0.07 \\
		&1998 May 10 &N &1.31 \sig 0.02 &149.3 \sig 0.4 &1.74 \sig 0.05 &\nodata &\nodata &\nodata \\
		&1998 May 10 &IHW18 &1.30 \sig 0.02 &149.1 \sig 0.4 &1.29 \sig 0.02 &\nodata &\nodata &\nodata \\
		&1999 May 05 &N &1.32 \sig 0.02 &149.4 \sig 0.4 &2.51 \sig 0.06 &2.63 \sig 0.07 & 3.63 \sig 0.05 & 2.26 \sig 0.07 \\
		&1999 May 05 &IHW18 &1.30 \sig 0.04 &149.2 \sig 0.6 &1.81 \sig 0.06 &1.11 \sig 0.11 &1.75 \sig 0.15 &0.63 \sig 0.13 \\
HBC 679    	&2001 May 26 &L &4.33 \sig 0.18 &57 \sig 1 &4.76 \sig 0.7 &8.64 \sig 0.02 &10.34 \sig 0.15 &8.43 \sig 0.04 \\
                &2001 May 10 &8.8 &\nodata &\nodata &\nodata &\nodata &\nodata &\nodata \\
	        &2001 May 10 &12.5 &\nodata &\nodata &\nodata &\nodata &\nodata &\nodata \\
\enddata
\\
\tablenotetext{a}{For the filters used here, vega has the following fluxes: N band
- 33.99 Jy; IHW18 - 11.96 Jy; 8.8\,\micron\, - 49.26 Jy; SiC - 28.24
Jy; 12.5\,\micron\, - 24.96 Jy. See Appendix \ref{calib} for more
details on the absolute photometric calibration.}
\tablenotetext{b}{Represents the magnitude of both primary and secondary combined, unless the 
value is starred. In this case, the observations reveal one point source and it is 
not clear which component the flux belongs to.}
\tablenotetext{c}{Chopping errors provide unresolved flux only.}
\tablenotetext{d}{Secondary marginally resolved in IRTF $L$-band observations causing large uncertainties 
in fluxes from PSF fitting. Used $L$-band results from \citet{wg01} for final sample.}
\tablenotetext{e}{A third object, situated 11\farcs1 from NTTS155203-2338A at a position angle of 164\fdg4,
was observed at K band (12.2 \sig 0.2 mag) and L band (11.57 \sig 0.01 mag).
Given the preponderance of objects within 1\arcmin of NTTS 155203-2338, this is likely to be a
background object but has not been confirmed.}
\tablenotetext{f}{A third object, located 16\farcs5 away from DoAr 26 A at a position angle of 49\fdg3, 
was also detected, with K=11.6 \sig 0.2 mag and L=11.6 \sig 0.8 mag. It is not known 
whether this object is physically associated with DoAr 26 or not.}
\end{deluxetable}

\clearpage

\begin{deluxetable}{llcccccccccccc}
\tabletypesize{\scriptsize}
\setlength{\tabcolsep}{0.02in}
\tablewidth{0pt}
\tablecaption{Stellar Component Colors \label{colors}}
\tablehead{
\multicolumn{1}{l}{Object} &\multicolumn{1}{c}{Component} &\multicolumn{1}{c}{$A_{V}$\tablenotemark{a}} 
&\multicolumn{1}{c}{SpTy} &\multicolumn{2}{c}{Dereddened} 
&\multicolumn{2}{c}{Ref\tablenotemark{b}} 
&\multicolumn{1}{c}{line?} &\multicolumn{1}{c}{EWH$\alpha$}
&\multicolumn{2}{c}{TT} &\multicolumn{1}{c}{Ref\tablenotemark{b}} &\multicolumn{1}{c}{notes} \\
\multicolumn{1}{l}{} &
\multicolumn{1}{l}{} &
\multicolumn{1}{c}{} &
\multicolumn{1}{c}{} &
\multicolumn{1}{c}{$K-L$} &
\multicolumn{1}{c}{$K-[N]$} &
\multicolumn{1}{c}{MIR} &
\multicolumn{1}{c}{NIR} &
\multicolumn{1}{c}{} &
\multicolumn{1}{c}{\AA} &
\multicolumn{1}{c}{class} &
\multicolumn{1}{c}{type} &
\multicolumn{1}{}{line} &
\multicolumn{1}{c}{}
}
\startdata
GG Tau &Aa  &0.75 &K7 &0.79 \sig 0.15 &2.64 \sig 0.17 &1 &3.2 &\Ha &42 &II &C &5 & \\
       &Ab  & &M0.5 &0.73 \sig 0.16 &3.42 \sig 0.14 &1 &3.2 &\Ha &21 &II &C &5 & \\
Haro 6-37 &B  &3.89 &M1 &0.58 \sig 0.08 &3.76 \sig 0.15 &1 &2 &\Ha &184 &II &C &2 & \\
GH Tau 	&A   &1.04 &M2 &0.64 \sig 0.25 &2.83 \sig 0.26 &1 &2 &\Ha &10 &II &C &5 & \\
 	&B   &  &M2 &0.81 \sig 0.23 &2.23 \sig 0.23 &1 &2 &\Ha &10 &II &C &5 & \\
LkHa 332 &A  &3.58 &K7 &0.87 \sig 0.10 &3.06 \sig 0.17 &1 &3.2 &\Ha &10.9 &II &C &5 & \\
	 &B  & &M2.5 &-0.16 \sig 0.14 &3.29 \sig 0.16 &1 &3.2 &\Ha &6.2 &II &W &5 &passive \\
UZ Tau E &Aa-Ab   &7.13 &M1 &0.44 \sig 0.2 &3.67 \sig 0.13 &1 &1 &\Ha &74. &II &C &2 & \\
UZ Tau W &A     &0.28 &M2 &1.00 \sig 0.60 &2.25 \sig 0.15 &1 &2.1 &\Ha &54. &II &C &5 & \\
  	 &B    & &M3 &1.05 \sig 0.61 &2.68 \sig 0.18 &1 &2.1 &\Ha &97. &II &C &5 & \\
DD Tau 	&A     &1.61 &M3 &1.16 \sig 0.25 &3.79 \sig 0.14 &1 &2 &\Ha &206 &II &C &5 & \\
	&B     & &M3 &1.05 \sig 0.29 &3.66 \sig 0.20 &1 &2 &\Ha &635. &II &C &5 & \\
Haro 6-28 &A &3.09 &M2 &0.82 \sig 0.30 &2.23 \sig 0.17 &1 &3.1 &\Ha &57 &II &C &5 & \\
	  &B & &M3.5 &0.53 \sig 0.32 &2.80 \sig 0.19 &1 &3.1 &\Ha &124 &II &C &5 & \\
FV Tau 	&A    &6.68 &K5 &0.27 \sig 0.21 &3.47 \sig 0.19 &1 &3.2 &\Ha &15. &II &C &5 & \\
	&B    &     &K6 &0.95 \sig 0.22 &3.01 \sig 0.17 &1 &3.2 &\Ha &63. &II &C &5 & \\
T Tau 	&A     &[0.0] &K1 &1.05 \sig 0.32 &3.84 \sig 0.30 &1 &2 &\Ha &60 &II &C &2 & \\
 	&Ba-Bb     & &\nodata &3.21 \sig 0.32 &5.78 \sig 0.31 &1 &2 &\nodata &\nodata &I &C &2 &IRC \\
FV Tau/c &A  &5.23 &M2.5 &0.17 \sig 0.05 &2.29 \sig 0.08 &1 &3.2 &\Ha &17 &II &W &2,5 &passive \\
	 &B  & &M3.5 &1.81 \sig 0.08 &5.40 \sig 0.10 &1 &3.2 &\Ha &224 &I &C &2,5 &IRC \\
FQ Tau 	&A    &0.35 &M3 &0.59 \sig 0.07 &2.95 \sig 0.11 &1 &1 &\Ha &110 &II &C &5 & \\
	&B    & &M3.5 &0.73 \sig 0.06 &2.61 \sig 0.11 &1 &1 &\Ha &23 &II &W\tablenotemark{c} &5 &passive \\
UY Aur 	&A    &2.05 &K7 &0.94 \sig 0.06 &4.16 \sig 0.10 &1 &2 &\Ha &36. &II &C &5 & \\
	&B    & &M2 &0.89 \sig 0.07 &4.13 \sig 0.10 &1 &2 &\Ha &59. &II &C &5 & \\
FX Tau 	&A    &2.24 &M1 &0.87 \sig 0.15 &3.09 \sig 0.18 &1 &2.1 &\Ha &13 &II &C &5 & \\
	&B    & &M4 &0.55 \sig 0.17 &2.72 \sig 0.17 &1 &2.1 &\Ha &1 &II &W &5 &passive \\
RW Aur 	&A    &[0.0] &K1 &1.11 \sig 0.22 &3.93 \sig 0.22 &1 &3.2 &\Ha &76 &II &C &2 &\\
	&B    & &K5 &0.73 \sig 0.25 &2.69 \sig 0.25 &1 &3.2 &\Ha &43. &II &C &2 &\\
GG Tau  &Ba   &[0.0] &M6 &0.81 \sig 0.04 &3.39 \sig 0.11 &1 &2 &\Ha &21 &II &C &2 &\\
 	&Bb   & &M7.5 &0.79 \sig 0.17 &3.22 \sig 0.13 &1 &2 &\Ha &32. &II &C &2 &\\
HK Tau 	&A    &3.41 &M1 &0.64 \sig 0.04 &3.14 \sig 0.08 &14 &2 &\Ha &50. &II &C &4 & \\
 	&B    & &M2 &0.28 \sig 0.12 &2.77 \sig 0.17 &14 &2 &\Ha &12.5 &II &C &4 & \\
IT Tau 	&A    &1.87 &K3 &0.24 \sig 0.09 &2.59 \sig 0.12 &1 &2.1 &\Ha &21.7 &II &C &4 & \\
	&B    & &M4 &0.77 \sig 0.13 &3.42 \sig 0.16 &1 &2.1 &\Ha &147. &II &C &4 &  \\
DK Tau 	&A    &0.35 &K9 &0.97 \sig 0.04 &3.99 \sig 0.14 &1 &2 &\Ha &31 &II &C &4 & \\
	&B    & &M1 &1.02 \sig 0.08 &3.22 \sig 0.15 &1 &2 &\Ha &118. &II &C &4 & \\
UX Tau 	&A   &[0.0] &K5 &0.76 \sig 0.06 &1.88 \sig 0.12 &1 &2 &\Ha &9.5 &II &C &2 & \\
	&B   & &M2 &0.09 \sig 0.06 &0.67 \sig 0.13 &1 &2 &\Ha &4.5 &III &W &2 & \\	
	&C   & &M5 &0.38 \sig 0.12 &0.51 \sig 0.11 &1 &2 &\Ha &8.5 &III &W &2 & \\
V710 Tau &A   &[0.0] &M0.5 &0.44 \sig 0.04 &3.32 \sig 0.11 &1 &2 &\Ha &69 &II &C &2 & \\
	&B   & &M2.5 &0.31 \sig 0.04 &0.29 \sig 0.11 &1 &2 &\Ha &7.2 &III &W &2 & \\
HV Tau 	&A    &2.42 &M2 &0.07 \sig 0.06 &0.13 \sig 0.14 &1 &1 &\Ha &4.3 &III &W &2 & \\
	&B    & &\nodata &0.73 \sig 0.11 &4.27 \sig 0.17 &1 &1 &\Ha &15. &II &C &2 & \\	
NTTS155203-2338	&A &[0.0] &\nodata &0.52 \sig 0.04 &0.11 \sig 0.06 &1 &1 &\brg &\nodata &III &W &10 & \\
		&B & &\nodata &0.08 \sig 0.09 &-0.27 \sig 0.05 &1 &1 &\brg &\nodata &III &W &10 & \\
WSB 18 	&A         &2.66 &M2 &0.15 \sig 0.24 &0.86 \sig 0.21 &1 &1 &\Ha &8.4 &III &W &9 & \\
	&B         & &M2.5 &0.49 \sig 0.09 &2.48 \sig 0.13 &1 &1 &\Ha &140 &II &C &9 & \\
WSB 26 	&A         &0.79 &M0 &0.70 \sig 0.09 &2.57 \sig 0.14 &1 &1 &\Ha &109 &II &C &9 & \\
	&B         & &M3 &0.64 \sig 0.08 &3.77 \sig 0.14 &1 &1 &\Ha &178 &II &C &9 & \\
ROX 15\tablenotemark{d} &A &8.43 &M3 &0.37 \sig 0.12 &2.00 \sig 0.10 &1 &1 &\brg &\nodata &II &C &8,10 &\\
	&B	 & &M3 &0.47 \sig 0.08 &1.75 \sig 0.12 &1 &1 &\brg &\nodata &II &C &8,10 & \\
AS 205 	&A         &2.41 &K5 &0.92 \sig 0.15 &3.35 \sig 0.18 &1 &1 &\Ha &220 &II &C & & \\
	&B         & &M3 &0.88 \sig 0.13 &3.37 \sig 0.19 &1 &1 &\Ha &55 &II &C & & \\
WSB 19 	&A         &1.68 &M3 &0.62 \sig 0.05 &4.02 \sig 0.07 &1 &1 &\Ha &56 &II &C &9 & \\
	&B         & &M5 &0.36 \sig 0.10 &3.89 \sig 0.12 &1 &1 &\Ha &37 &II &C &9 & \\
DoAr 24E &A       &4.75 &K5 &0.45 \sig 0.17 &2.63 \sig 0.18 &13 &1 &\brg &\nodata &II &C &8,10,15 & \\
	&B       & &\nodata &1.62 \sig 0.17 &5.05 \sig 0.15 &13 &1 &\brg &\nodata &I &C &8,10 &IRC \\
DoAr 26	&A       &1.64 &M4 &0.68 \sig 0.26 &2.41 \sig 0.26 &1 &1 &\brg &\nodata &II &C &8,10 & \\
	&B       &  &M6 &0.10 \sig 0.22 &1.88 \sig 0.22 &1 &1 &\brg &\nodata &II &W &8,10 &passive \\
WSB 28 	&A    	&3.28 &M3 &-0.13 \sig 0.13 &-0.07 \sig 0.11 &1 &1 &\brg &\nodata &III &W &8 & \\
	&B        & &M7 &0.02 \sig 0.44 &2.24 \sig 0.36 &1 &1 &\brg &\nodata &II &W &8 &passive \\
S CrA 	&A          &0.41 &K3 &1.40 \sig 0.11 &3.39 \sig 0.12 &1 &1 &\brg &\nodata &II &C &8 & \\
	&B          & &M0 &1.33 \sig 0.13 &3.67 \sig 0.14 &1 &1 &\brg &\nodata &II &C &8 & \\
LkCa 3 	&A     &1.20 &\nodata &-0.02 \sig 0.04 &0.26 \sig 0.11 &1 &1 &\nodata &\nodata &III & & & \\
 	&B     & &\nodata &-0.01 \sig 0.05 &0.07 \sig 0.08 &1 &1 &\nodata &\nodata &III & & & \\
HBC 411 &A   &6.71 &\nodata &0.65 \sig 0.05 &3.41 \sig 0.13 &1 &2 &\nodata &\nodata  &II & & &\\
	&B   & &\nodata &0.45 \sig 0.09 &2.36 \sig 0.13 &1 &2 &\nodata &\nodata &II & & & \\
V853 Oph &A 	&0.29 &\nodata &0.56 \sig 0.04 &2.26 \sig 0.10 &1 &1 &\nodata &\nodata &II & & & \\
	&B	& &\nodata &0.46 \sig 0.07 &2.17 \sig 0.07 &1 &1 &\nodata &\nodata &II  & & & \\
SR 9 	&A          &[0.0] &\nodata &0.62 \sig 0.03 &2.81 \sig 0.10 &1 &1 &\Ha &10.5 &II & & & \\
	&B          & &\nodata &0.36 \sig 0.07 &2.95 \sig 0.12 &1 &1 &\Ha &13.8 &II & & & \\
\enddata
\tablenotetext{a}{$A_{V}$ is measured from photometry for combined system. Values given as [0.0]
indicate an assumed extinction of 0.0 mag in cases where J-H and H-K colors fall below the CTTS
locus.}
\tablenotetext{b}{References for $K-L$, $K-N$ color, line emission and
spectral type are
1 - This work; 2 - \citet{wg01}; 3 - Woitas et al. 2001b; 
4 - \citet{duchene99}; 5 - \citet{hk03}; 6 - \citet{hss94}; 
7 - \citet{kenyon98}; 8 - \citet{pgs03};
9 - \citet{bz97}; 10 - L. Prato, priv. comm; 11 - \citet{martin98}; 
12 - \citet{hkl}; 13 - \citet{stanke};
14 - \citet{mccabe03}; 15 - \citet{doppmann03}. If a 
reference is given as a fractional number, the unitary number reference
corresponds to the $K$ band and the fractional part corresponds to
the $L$-band.}
\tablenotetext{c}{Hartigan \& Kenyon (2003) find an EW(H$\alpha$) = 23 $\pm$ 4 \AA, 
on the border between CTTS/WTTS. Based on this, the lack of veiling and 
[O\,I] emission, and NIR excess, they define it to be a WTTS.
We use their classification here.}
\tablenotetext{d}{ROX 15 flickers in accretion; Prato et al. (2003)
see no \brg in either component, whereas Greene \& Meyer (1995) and
Luhman \& Rieke (1999) do see \brg in resolved system. Went with C/C
because system seen to be accreting at least on some occasions.}
\end{deluxetable}
\end{landscape}

\clearpage

\begin{deluxetable}{lccccc}
\tabletypesize{\footnotesize}
\tablewidth{0pt}
\tablecaption{Flux Densities of Calibrators\label{primphot}}
\tablehead{
\multicolumn{1}{l}{Name} &\multicolumn{2}{c}{OSCIR (Jy)} &\multicolumn{3}{c}{LWS (Jy)}\\
\cline{4-6}
\multicolumn{1}{l}{} &\multicolumn{1}{c}{N} &\multicolumn{1}{c}{IHW18} &\multicolumn{1}{c}{8.8\micron} &\multicolumn{1}{c}{12.5\micron} &\multicolumn{1}{c}{SiC} 
}
\startdata
{\it primary calibrators:} & & & & & \\
$\alpha$ Lyr &33.99 &11.96 &49.26 &24.96 &28.24 \\
$\rho$ Boo &20.81 &7.52 &28.23 &15.72 &17.8 \\
$\gamma$ Aql &66.11 &24.29 &89.2 &50.47 &56.78 \\
$\alpha$ Boo &619.12 &219.21 &877.25 &457.78 &520.11 \\
$\alpha$ Ari &68.8 &24.93 &95.37 &51.72 &58.52 \\
$\beta$ Peg &323.50 &122.23 &439.96 &246.84 & 278.75 \\
10 And &5.07 &1.87 &6.73 &3.91 &4.42 \\
HD 27482 &5.95 &2.33 &7.86 &4.78 &5.25 \\
HD 1240 &8.35 &3.08 &11.08 &6.44 &7.27 \\
{\it secondary calibrators:} & & & & & \\
$\alpha$ CrB &4.24 \sig 0.25 &1.35 \sig 0.01 &6.36 \sig 0.6 &3.3 \sig 0.2 & \\ 
$\alpha$ CMi &69.73 \sig 5.38 &23.1 \sig 0.6 & & &53.88 \sig 0.58 \\
$\alpha$ Sco &[2269] &[815.75] & & & \\
$\delta$ Sco &6.09 \sig 0.9 & 2.89\sig 0.3 & & & \\
$\mu$ UMa &[96.46] &[34.02] & & &  \\
SAO90816 &11.03 \sig 0.9 &3.56 \sig 0.1 & & & \\
\enddata
\end{deluxetable}

\clearpage

\begin{deluxetable}{lcccccccccccc}
\tabletypesize{\scriptsize}
\tablewidth{0pt}
\tablecaption{Single Stars \label{singles}}
\tablehead{
\multicolumn{1}{l}{Star} &\multicolumn{1}{c}{SpTy} &\multicolumn{1}{c}{$A_{v}$} &\multicolumn{3}{c}{Observed\tablenotemark{a}} &\multicolumn{2}{c}{EW\Ha} &\multicolumn{1}{c}{Multiplicity\tablenotemark{b}} 
&\multicolumn{2}{c}{TT} &\multicolumn{1}{c}{Obs.} &\multicolumn{1}{c}{Notes}\\
\cline{4-6}
\colhead{} &
\colhead{} &
\colhead{} &
\colhead{$K-L$} &
\colhead{$K-[N]$} &
\colhead{ref\tablenotemark{b}} &
\colhead{(\AA)} &
\colhead{ref\tablenotemark{b}} &
\colhead{reference} &
\colhead{class} &
\colhead{type} &
\colhead{method} &
\colhead{}
}
\startdata
GM Aur &K3 &0.25 &0.33 \sig 0.06 &3.63  &1 &96 &5 &10 &II &C &speckle& \\
CW Tau &K3 &1.22 &1.20 \sig 0.24 &3.33 \sig 0.19 &1 &135 &5 &9,10 &II &C &speckle & \\
DS Tau &K5 &0.00 &0.71 \sig 0.08 &3.13  &1 &59 &5 &10 &II &C &speckle & \\
LkCa 15 &K5 &0.41 &0.57 \sig 0.06 &2.48  &1 &13 &5 &10 &II &C &speckle & \\
BP Tau &K7 &0.72 &0.57 \sig 0.23 &2.93 \sig 0.29 &1 &40 &6 &9 &II &C &speckle& \\
GK Tau &K7 &0.32 &0.85 \sig 0.07 &3.21 \sig 0.21 &1 &22 &7 &9 &II &C &speckle& \\ 
AA Tau &K7 &0.72 &0.90 \sig 0.12 &3.12 \sig 0.19 &1 &37 &5 &10 &II &C &speckle& \\
DL Tau &K7 &0.00 &1.21 \sig 0.34 &3.43 \sig 0.20 &1 &105 &5 &10 &II &C &speckle & \\
GI Tau &K7 &0.93 &0.96 \sig 0.18 &3.74 \sig 0.08 &1 &20 &7 &9,10 &II &C &speckle& \\
CI Tau &K7 &1.68 &0.90 \sig 0.17 &3.36  &1 &102 &5 &9,10 &II &C &speckle& \\
DN Tau &M0 &0.95 &0.55 \sig 0.22 &2.53 \sig 0.11 &1 &12 &5 &10 &II &C &speckle& \\
IP Tau &M0 &0.37 &0.89  &2.80  &1 &11 &5 &10 &II &C &speckle& \\
DO Tau &M0 &1.52 &1.21 \sig 0.19 &3.85 \sig 0.17 &1 &100 &5 &9,10 &II &C &speckle& \\
DP Tau &M0.5 &2.63 &0.95 \sig 0.29 &3.80  &1 &85 &5 &10 &II &C &speckle& \\
DE Tau &M1 &1.18 &0.97 \sig 0.35 &2.85 \sig 0.13 &1 &54 &6 &9 &II &C &speckle& \\
DM Tau &M1 &0.25 &0.83 \sig 0.18 &4.45  &1 &139 &5 &10 &II &C &speckle& \\
FM Tau &M1 &0.33 &0.70 \sig 0.24 &2.53  &1 &62 &7 &10 &II &C &speckle& \\
DH Tau &M2 &0.25 &0.64 \sig 0.05 &2.44 \sig 0.79 &1 &72 &7 &9 &II &C &speckle& \\%
V827 Tau &K7 &0.00 &0.10 \sig 0.18 &0.31  &1 &1.8 &5 &10 &III &W &speckle& \\
LkCa 4 &K7 &1.03 &0.18  &0.08  &1 &5 &6 &10 &III &W &speckle & \\
HBC 374 &K7 &2.45 &0.01 \sig 0.16 &0.30  &1 &3 &6 &9,10 &III &W &speckle & \\
V819 Tau &K7 &2.52 &0.21  &0.61  &1 &1.7 &6 &9,10 &III &W &speckle& \\
IQ Tau &M0.5 &1.02 &0.96  &3.1  &1 &8 &5 &9,10 &II &W &speckle &passive \\
LkCa 1 &M4 &1.30 &0.07  &0.54  &1 &4 &5 &10 &III &W &speckle& \\
SR 4 &K5 &1.25 &0.84  &3.29  &2,3 &84 &5 &9,13 &II &C &speckle/lunar& \\
ROX 7 &K7 &6.23 &0.29  &$<$1.59  &3 &1.5 &8 &11 &- &W &speckle & \\
ROX 44 &K3 &0.90 &0.77  &2.6  &4 &54 &5 &12,13 &II &C &speckle/lunar & \\
\enddata
\tablenotetext{a}{Uncertainties in the observed colors 
are given when available in the literature.}
\tablenotetext{b}{Source: 
[1] - \citet{kh95}
[2] - \citet{myers87}
[3] - \citet{wly89}
[4] - \citet{rss76}
[5] - \citet{hbc}
[6] - \citet{ss94}
[7] - \citet{hss94}
[8] - \citet{ba92}
[9] - \citet{ghez93}; 
[10] - \citet{leinert93};
[11] - \citet{ageorge97};
[12] - \citet{barsony03};
[13] - \citet{simon95}
}
\end{deluxetable}

\clearpage

\figcaption[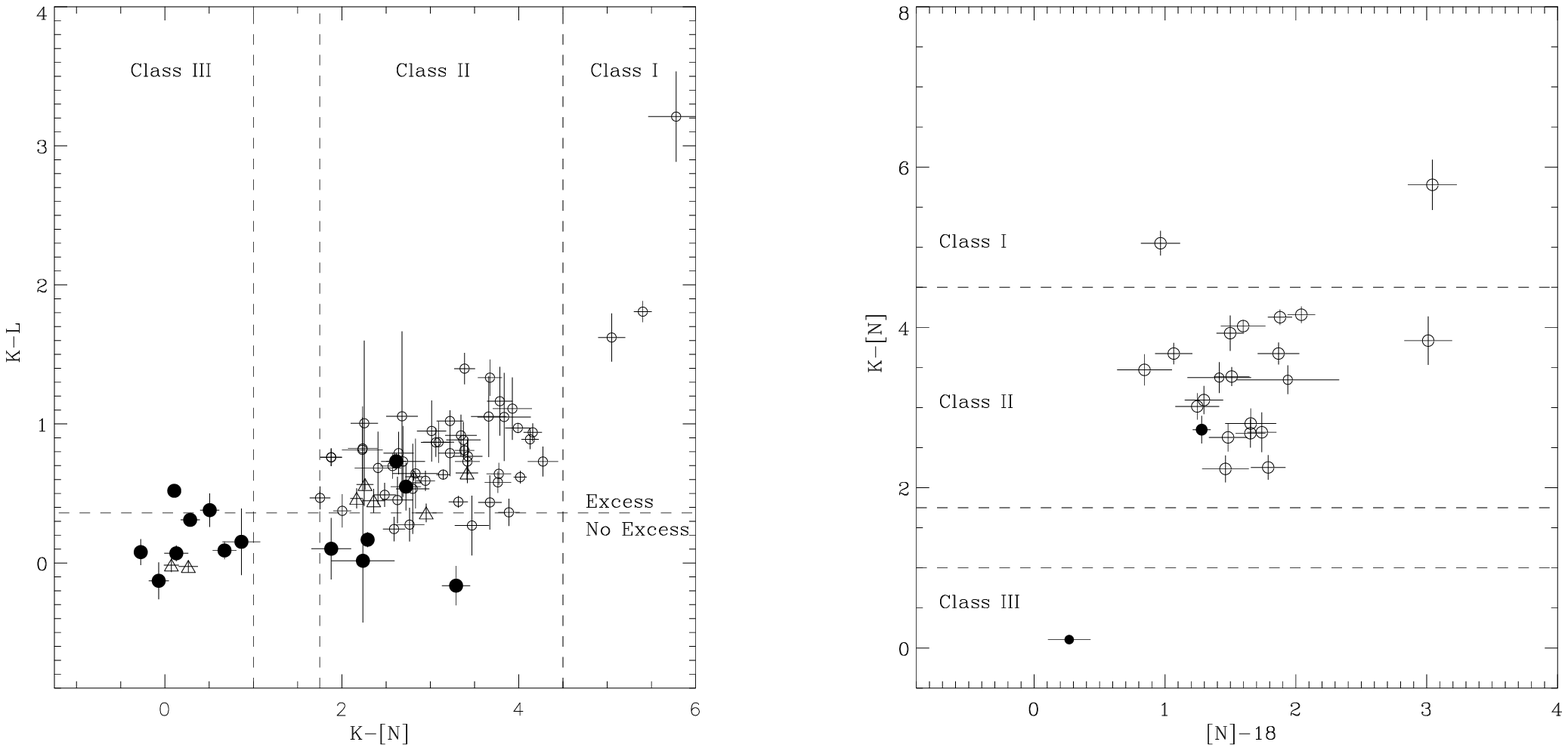]{Infrared
colors for our sample. {\it(left):} The de-reddened \kl vs. \knobs
color plane.  The symbol type represents the accretion status:
actively accreting (unfilled circle), not accreting (filled circle),
or accretion status unknown (triangle). The \knobs color definitions
for the different T Tauri classes are also shown, along with the \kl
color limit used to define whether the de-reddened color is in excess
of that expected from a photosphere. Of particular note is that 10\%
of the class II stars are not accreting (i.e., are passive disks) and
these systems, on average, have lower \kl colors than accreting class
II stars. {\it (right):} The de-reddened K-[N] vs. [N]-18 color plane,
with symbol types same as in left panel. \label{colorplane_dered}}

\figcaption[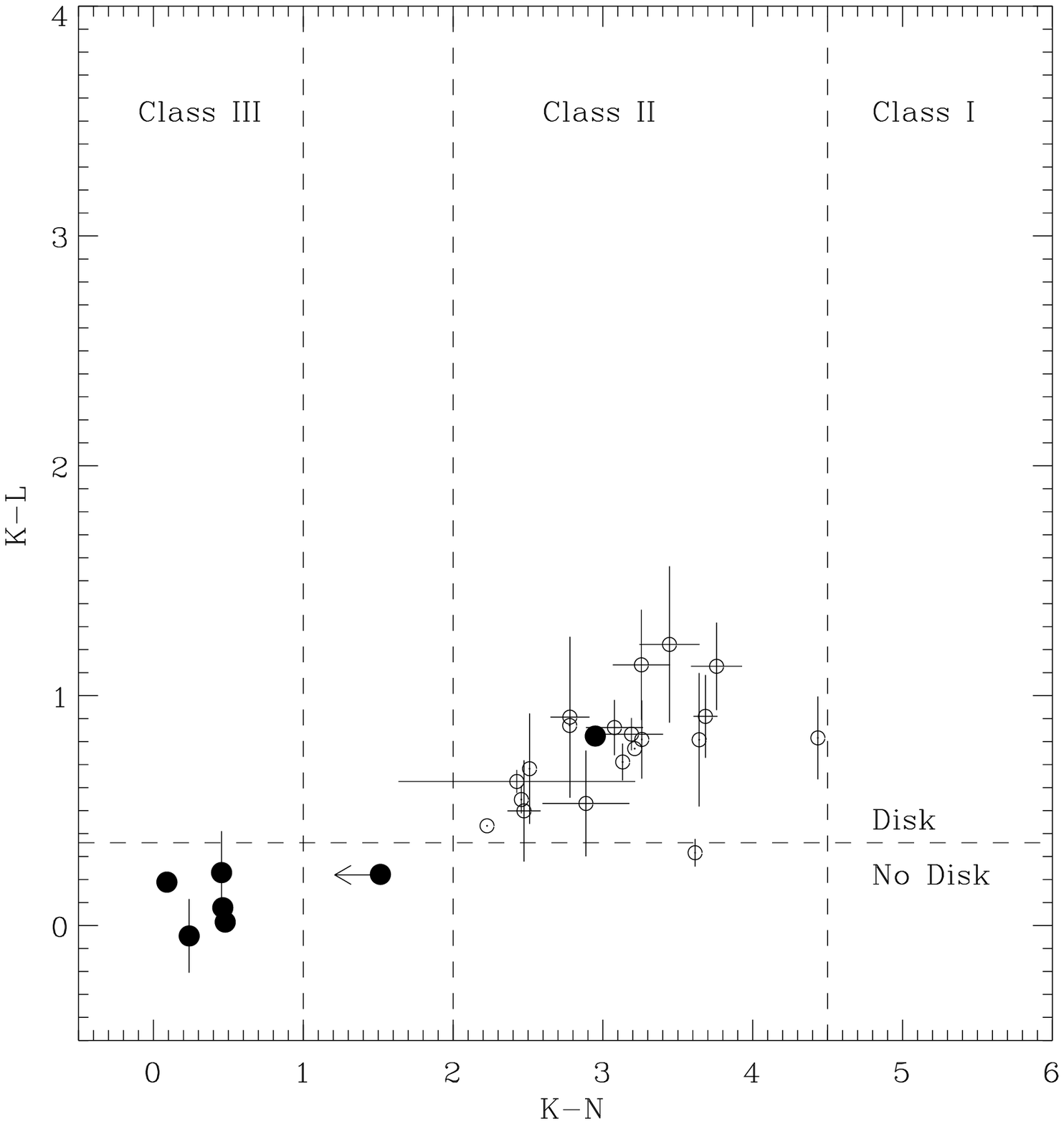]{ The
\kl vs \knobs de-reddened color plane for a sample of single stars
from \citet{wg01} and \citet{barsony03}. Unfilled circles represent
CTTS whereas filled circles symbols represent WTTS.  Of the 20 class
II stars in the sample, 1 is not accreting, corresponding to a 5\%
detection rate of passive disks in single stars, similar to that seen
in binary systems. \label{colorsingles}}

\figcaption[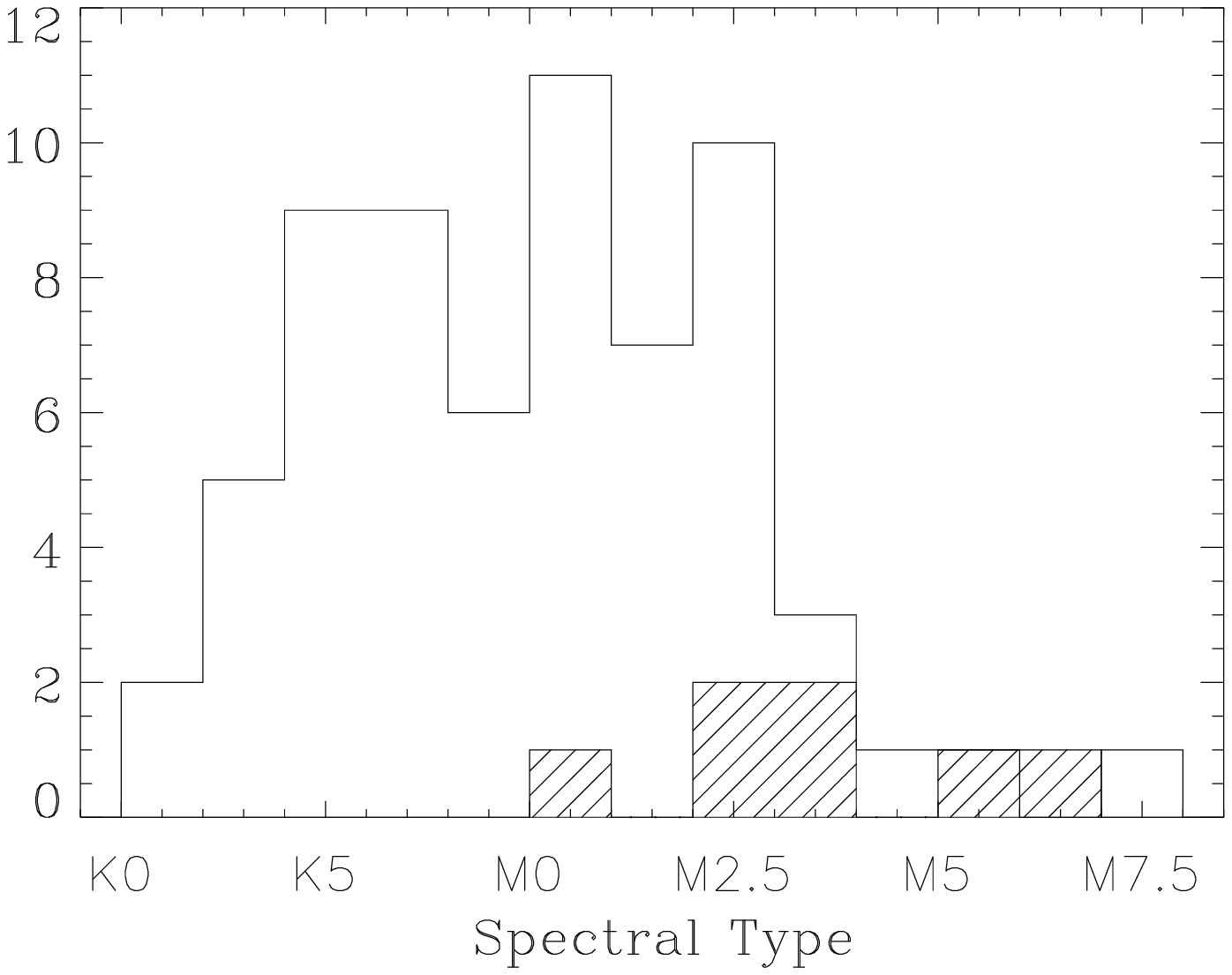]{A histogram
of the spectral types for all class II-CTTS stars in both our binary star
sample and single star sample is shown. For comparison, the
distribution of spectral types for the passive disk stars is
overplotted (hatch-marked). All passive disks are found around M type
stars and a K-S test demonstrates that these two distributions are
unlikely to have been drawn from the same parent population. \label{hist_spty}}

\clearpage
\begin{figure}
\plotone{f1.eps}
\end{figure}

\begin{figure}
\plotone{f2.eps}
\end{figure}

\begin{figure}
\plotone{f3.eps}
\end{figure}

\end{document}